\documentclass{JHEP3}

\preprint{LU-TP 05-11\\
  hep-ph/0503293
}

\usepackage{cite}
\usepackage{epsfig}
\usepackage{xspace}
\usepackage{graphics}
\usepackage[latin1]{inputenc}

\def\hide#1{}

\newcommand{\madgraph}{M\scalebox{0.8}{AD}G\scalebox{0.8}{RAPH}\xspace}
\newcommand{\madevent}{M\scalebox{0.8}{AD}E\scalebox{0.8}{VENT}\xspace}
\newcommand{\ktclus}{KTC\scalebox{0.8}{LUS}\xspace}

\newcommand{\apacic}{\scalebox{0.8}{APACIC++}\xspace}
\newcommand{\herwig}{\scalebox{0.8}{HERWIG}\xspace}

\newcommand{\sherpa}{S\scalebox{0.8}{HERPA}\xspace}
\newcommand{\diclus}{D\scalebox{0.8}{ICLUS}\xspace}
\newcommand{\pythia}{P\scalebox{0.8}{YTHIA}\xspace}

\newcommand{\ariadne}{A\scalebox{0.8}{RIADNE}\xspace}

\newcommand{\as}{\ensuremath{\alpha_{\mathrm{s}}}}

\newcommand{\kT}{\ensuremath{k_{\perp}}}
\newcommand{\pT}{\ensuremath{p_{\perp}}}

\newcommand{\kTi}[1]{\ensuremath{k_{\perp #1}}}
\newcommand{\pTi}[1]{\ensuremath{p_{\perp #1}}}

\newcommand{\ltaeq}{\raisebox{-0.8mm}%
{\hspace{1mm}$\stackrel{<}{\sim}$\hspace{1mm}}}

\newcommand{\particle}[1]{\ensuremath{\mathrm{#1}}}
\newcommand{\antiparticle}[1]{\ensuremath{\bar{\mathrm{#1}}}}
\newcommand{\el}{\particle{e}}

\newcommand{\g}{\particle{g}}
\newcommand{\p}{\particle{p}}
\newcommand{\q}{\particle{q}}

\newcommand{\W}{\particle{W}}
\newcommand{\Wp}{\ensuremath{\W^+}}

\newcommand{\qu}{\particle{u}}
\newcommand{\qd}{\particle{d}}

\newcommand{\pbar}{\antiparticle{p}}
\newcommand{\qbar}{\antiparticle{q}}

\newcommand{\ubar}{\antiparticle{u}}
\newcommand{\dbar}{\antiparticle{d}}

\newcommand{\ee}{\ensuremath{\el^+\el^-}}
\newcommand{\tee}{\ensuremath{\el^+\el^-}\xspace}

\def\mrm#1{\mathrm{#1}}

\def\sub#1{\ensuremath{_{\mrm{#1}}}}
\def\sup#1{\ensuremath{^{\mrm{#1}}}}

\def\sud#1{\ensuremath{\Delta_{S_{#1}}}}

\def\f2d3{\ensuremath{F_2^{\mrm{D}3}}}

\def\done#1{}

\newcommand{\eqref}[1]{eq.~(\ref{#1})\xspace}
\newcommand{\eqsref}[1]{eqs.~(\ref{#1})\xspace}

\newcommand{\dash}{\ensuremath{\!\!-\!}}

\newcounter{aenumct}
\newenvironment{anumerate}{\begin{list}{(\alph{aenumct})}%
{\usecounter{aenumct}\setlength{\topsep}{1mm}%
\setlength{\partopsep}{1mm}\setlength{\itemsep}{0mm}%
\setlength{\parsep}{1mm}}}{\end{list}}

\renewenvironment{itemize}{\begin{list}{$\bullet$}%
{\setlength{\topsep}{0mm}\setlength{\partopsep}{1mm}%
\setlength{\itemsep}{0mm}\setlength{\parsep}{1mm}}}%
{\end{list}}

\newcounter{enumct}
\renewenvironment{enumerate}{\begin{list}{\arabic{enumct}.}%
{\usecounter{enumct}\setlength{\topsep}{1mm}%
\setlength{\partopsep}{1mm}\setlength{\itemsep}{0mm}%
\setlength{\parsep}{1mm}}}{\end{list}}

\usepackage{axodraw}

\skip\footins = 1\bigskipamount plus 2pt minus 4pt                              

\title{\boldmath$\W+$jets Matrix Elements and the Dipole Cascade}

\author{Nils Lavesson and Leif Lönnblad\\
  Dept.~of Theoretical Physics,
  Sölvegatan 14A, S-223 62  Lund, Sweden\\
  E-mail: \email{Nils.Lavesson@thep.lu.se}
    and \email{Leif.Lonnblad@thep.lu.se}}
  
  \abstract{We extend the algorithm for matching fixed-order
    tree-level matrix element generators with the Dipole Cascade Model
    in \ariadne to apply to processes with incoming hadrons. We test
    the algoritm on for the process \W+$n$ jets at the Tevatron, and
    find that the results are fairly insensitive to the cutoff used to
    regularize the soft and collinear divergencies in the tree-level
    matrix elements. We also investigate a few observables to check
    the sensitivity to the matrix element correction.}

\begin{document}
 
\sloppy
 
\section{Introduction}
\label{sect-intro}

Parton Shower based Monte Carlo Event Generators (PSEGs) have
developed into essential tools in High Energy Physics. Without them it
is questionable if it at all would be possible to embark on
large-scale experiments such as the LHC. Although they are based on
leading logarithmic approximations and phenomenological hadronization
models, they are typically able to describe hadronic final states in
great detail and, especially at LEP, with great precision. However,
there are problems. The description of final states which include more
than three hard jets is not very good, and when it comes to collisions
with incoming hadrons, the precision is generally lacking, especially
for small-$x$ processes. In this article we will address both these
problems.

The problem with describing several hard, well separated jets is
inherent in the leading log approximations, since they assume that
there is strong ordering between parton emissions, and hence only give
a good description of soft inter-jet and collinear intra-jet
emissions. Typically it is possible to correctly describe one
additional hard jet on top of the hard sub-process used as starting
point, by applying correction factors to the basic splitting
functions. However, to go beyond one additional jet is more difficult.

To describe events with several hard partons we can use so-called
Matrix Element Generators (MEGs), where the parton distributions can
be generated according to exact tree-level matrix elements.
Unfortunately these matrix elements are divergent in the soft and
collinear limits and a cutoff is needed to avoid these regions of
phase space. However, to generate realistic events we need to hadronize the
partons into jets of hadrons, and all reasonable hadronization models
require that also soft and collinear parton emissions are modeled
correctly. Hence the need for combining matrix element generators with
parton showers.

Combining these two approaches is, however, not trivial. The matrix
elements describe inclusive events, ie.\ events with at
least $n$ partons above some cutoff, while parton showers are
exclusive and describes events with exactly $n$ partons. Naively
adding parton showers to events generated by a MEG will therefore give
a very strong dependence on the cutoff used in the MEG, even if great
care is taken to avoid double-counting by only adding parton showers
below that cutoff.

A solution for this problem was presented by Catani et al.\ in
\cite{Catani:2001cc}. The procedure, generally referred to as CKKW,
relies on applying a jet clustering algorithm to partonic event from a
MEG generating zero, one, two, etc.\ additional hard jets above some
cutoff according to exact tree-level matrix elements. The repeated
clustering of two jets into one is then used to construct an ordered
set of scales corresponding to consecutive parton emissions. These
scales are used to calculate Sudakov form factors corresponding
to no-emission probabilities, which are used to reweight the MEG
events to make them exclusive. A parton shower can then be added with
a special veto to avoid double-counting of emissions above the cutoff.
In this way it was shown that the dependence on the cutoff cancels to
next-to-leading order accuracy.  The dependence was, however, still
quite visible, giving rise to annoying discontinuities in some
observables.

The basic CKKW prescription was improved\cite{Lonnblad:2001iq} when
implemented for the Dipole Cascade
model\cite{Gustafson:1988rq,Gustafson:1986db} in the \ariadne
program\cite{Lonnblad:1992tz}. Rather than using a jet clustering
algorithm to construct a set of scales, the \ariadne procedure
involves constructing complete intermediate states corresponding to a
series of emissions which the dipole cascade could have used to
produce a given state obtained from a MEG. A special veto algorithm is
used to calculate exactly the Sudakov form factors the cascade would
have used to produce the state, which are then used for reweighting.
Together with a special treatment of the MEG-produced states with
highest multiplicity, this procedure basically removes any visible
discontinuities due to the cutoff.

Both these procedures were originally developed for \tee. Recently
there has been some developments in applying them to hadronic
collisions, in particular for the \W$+$jets process, by Krauss et al.\ 
\cite{Krauss:2002up,Krauss:2004bs,Krauss:2005nu} and Mrenna and
Richardson \cite{Mrenna:2003if}.  An alternative procedure has also
been developed by Mangano\cite{MLM}, which is similar in spirit to
CKKW, but which has a simpler interface between the MEG and PSEG. This
development is very important for the LHC, where \W$+$jets is an
important background for almost any signal of new physics.

In this article we describe the extension of the CKKW procedure for
the dipole cascade in \ariadne to handle hadronic collisions, again
concentrating on the \W$+$jets process. The goal is to obtain a
procedure which gives as small cutoff dependence as was achieved for
\tee. However, we also expect to see differences w.r.t.\ the
procedures of Mrenna, Richardson and Krauss, since the dipole cascade
model for collisions with incoming
hadrons\cite{Andersson:1989gp,Lonnblad:1996ex} is different. The
standard initial-state parton shower approaches, such as those
implemented in \pythia\cite{Sjostrand:2001yu,Bengtsson:1987rw},
\herwig\cite{Corcella:2000bw} and
\sherpa/\apacic\cite{Gleisberg:2003xi,Krauss:2005re}, as
well as the Sudakov form factors used in CKKW, corresponds to a
DGLAP-resummation
\cite{Gribov:1972ri,Lipatov:1975qm,Altarelli:1977zs,Dokshitzer:1977sg}
of leading logarithms of the hard scale. In the dipole cascade,
however, also some terms corresponding to logarithms of $1/x$ are
resummed. Although not formally equivalent to neither BFKL
\cite{Kuraev:1976ge,Kuraev:1977fs, Balitsky:1978ic} or CCFM
\cite{Ciafaloni:1988ur,Catani:1990yc,Catani:1990sg,Marchesini:1995wr}
evolution, it has proven to be able to describe most features
small-$x$ final states at HERA, where all DGLAP-based parton showers
fail. Now, \W$+$jets is conventionally not considered to be a
small-$x$ process because of the hard scale, $m\sub{\W}$, being large,
but at the LHC the collision energies are so large there may be
substantial effects of terms proportional to
$\as^n\log(\sqrt{S}/m\sub{\W})^n$, where $m\sub{\W}/\sqrt{S}\sim x\sim
0.005$.

The layout of this article is as follows. We will first recap the main
points of the CKKW procedure and how it is implemented in \ariadne for
\tee in section \ref{sec:ckkw-dipole-cascade}, followed by a
description of how hadronic collisions are treated in \ariadne in
section \ref{sec:dipole-model-incom}. Then, in section
\ref{sec:prot-ckkw-boldm}, we will describe our CKKW implementation
for \W$+$jets, and hadronic collisions in general, starting with the
construction of intermediate states in \ref{sec:reconstr-emiss},
followed by the reweighting procedure in \ref{sec:reweighting-events}.
The results of our investigation of its performance are presented in
\ref{sec:results}.  Finally in section \ref{sec:conclusions} we
present our conclusions.

\section{CKKW and the Dipole Cascade Model}
\label{sec:ckkw-dipole-cascade}

When generating events with a PSEG, the procedure is to start from a
primary hard sub-process, typically a $2\to 2$ process such as
$\ee\to\q\qbar$ or $\q\qbar\to \W\to\bar{\nu}e$, and then to let the
incoming and outgoing quarks and gluons evolve a parton cascade in an
iterative $1\to2$ branching procedure. The emissions are ordered
according to some evolution scale $\rho$, where the maximum scale, $\rho_0$ is
typically given by the hardest scale in the primary sub-process, and
the minimum is some cutoff scale of the order of one GeV, typically
tuned to match a particular hadronization model.

We can write the exclusive cross sections for in this way generating
0,~1,~2,~\ldots\ additional partons above the cutoff, $\rho_c$,
as
\begin{eqnarray}
  \label{eq:psexpansion}
  \sigma_{+0} &=& \sigma_0\,\sud{0}(\rho_0,\rho_c) \nonumber\\
  d\sigma_{+1} &=& \sigma_0\,\as(\rho_1)
  c\sup{PS}_{11}
  \sud{0}(\rho_0,\rho_1)\sud{1}(\rho_1,\rho_c)d\rho_1d\Omega_1 \nonumber\\
  d\sigma_{+2} &=& \sigma_0\,\as(\rho_1)\as(\rho_2)
  c\sup{PS}_{22}
  \sud{0}(\rho_0,\rho_1)\sud{1}(\rho_1,\rho_2)\sud{2}(\rho_2,\rho_c)
  d\rho_1d\Omega_1d\rho_2d\Omega_2\nonumber\\
  &\vdots&\nonumber\\
  d\sigma_{+n} &=& \sigma_0\,c\sup{PS}_{nn}\sud{n}(\rho_n,\rho_c)\prod_{i=1}^n\as(\rho_i)
  \sud{i-1}(\rho_{i-1},\rho_i)
  d\rho_id\Omega_i\nonumber\\
  &\vdots&
\end{eqnarray}
where the ordering is $\rho_0>\rho_1>\ldots>\rho_n>\rho_c$ and
$\Omega_i$ symbolizes the phase space variables defining the $i$th
emission in addition to $\rho_i$ (typically some momentum fraction,
$z_i$, and some azimuth angle $\phi_i$). $\sud{i}(\rho_i,\rho_{i+1})$
is here the so-called Sudakov form factors giving the probability that
no emissions occurred from the state with $i$ additional partons
between the scales $\rho_i$ and $\rho_{i+1}$. The coefficients
$c_{nn}\sup{PS}$ are basically products of splitting functions which
depends on $\rho_i$ and $\Omega_i$, and we assume an implicit sum over
all possible flavour combinations. As we will see below, the
$c_{nn}\sup{PS}$ may also include ratios of parton density functions
(PDFs) in the case of incoming hadrons.

The Sudakov form factors are formally resummations of virtual diagrams
to all orders and can, in principle, be calculated analytically. They
would then only depend on the limiting scales, as is done in the
standard CKKW procedure. However, when explicitly interpreted as a
no-emission probability in a PSEG, as is done eg.\ in \ariadne, it
basically depends on all momenta in the partonic state. The typical
form of the Sudakov is
\begin{equation}
  \label{eq:sud}
  \sud{}(\rho_i,\rho_{i+1})=\exp\left(-\int_{\rho_{i+1}}^{\rho_i}
    \frac{d\rho}{\rho}\as(\rho)\int dz P(z)\right),
\end{equation}
which, of course, can be expanded in a series in \as, and we can
rewrite the exclusive cross sections as
\begin{eqnarray}
  \label{eq:psexpansion2}
  \sigma_{+0} &=& \sigma_0\,
  (1+c_{01}\sup{PS}\as+c_{02}\sup{PS}\as^2+\ldots) \nonumber\\
  d\sigma_{+1} &=& \sigma_0\,\as
  c\sup{PS}_{11}
  (1+c_{12}\sup{PS}\as+c_{13}\sup{PS}\as^2+\ldots)d\rho_1d\Omega_1 \nonumber\\
  d\sigma_{+2} &=& \sigma_0\,\as^2
  c\sup{PS}_{22}
  (1+c_{23}\sup{PS}\as+c_{24}\sup{PS}\as^2+\ldots)
  d\rho_1d\Omega_1d\rho_2d\Omega_2\nonumber\\
  &\vdots&\nonumber\\
  d\sigma_{+n} &=& \sigma_0\,\as^nc\sup{PS}_{nn}
  (1+c_{n,n+1}\sup{PS}\as+c_{n,n+2}\sup{PS}\as^2+\ldots)
  \prod_{i=1}^n d\rho_id\Omega_i\nonumber\\
  &\vdots&
\end{eqnarray}
to emphasize the resummation aspect. We note that even though all the
coefficients $c_{ij}\sup{PS}$ are divergent in the soft and collinear
limit when $\rho_c\to0$, the resummation to all orders in the Sudakovs
gives a finite result for each of the cross sections. Also, when
integrated over the allowed phase space the cross section of the
primary sub process $\sigma_0$ is retained,
\begin{equation}
  \label{eq:sumxsec}
\sum_0^\infty\sigma_{+i}=\sigma_0 .
\end{equation}

In contrast a MEG will generate inclusive partonic states with the
cross sections for generating \emph{at least} 0,~1,~2,~\ldots
additional jets given by\newpage
\begin{eqnarray}
  \label{eq:meexpansion}
  \sigma_{+0} &=& \sigma_0\,\nonumber\\
  d\sigma_{+1} &=& \sigma_0\,\as
  c\sup{ME}_{11}
  d\mathbf{\Omega}_1 \nonumber\\
  d\sigma_{+2} &=& \sigma_0\,\as^2
  c\sup{ME}_{22}
  d\mathbf{\Omega}_1d\mathbf{\Omega}_2\nonumber\\
  &\vdots&\nonumber\\
  d\sigma_{+n} &=& \sigma_0\,\as^nc\sup{ME}_{nn}
  \prod_{i=1}^n d\mathbf{\Omega}_i\nonumber\\
  &\vdots&
\end{eqnarray}
where the $\mathbf{\Omega}_i$ symbolizes all phase space variables
defining the $i$th parton, and the coefficients $c\sup{ME}_{ii}$ are
calculated using the exact tree-level matrix elements (including PDFs
in the case of incoming hadrons). Clearly we can not simply add
these cross sections, especially since each of the coefficients are
divergent if the soft and collinear limits are not cut off properly.

The advantage of using a MEG is that the exact tree-level matrix
elements are used, which means that also states with several hard
partons are described correctly. This is not the case for the PSEG,
where the coefficients are given by products of splitting functions,
which is only a good approximation in the limit of strongly ordered
emissions. On the other hand a MEG will not correctly treat soft and
collinear partons, where the coefficients are large and need to be
resummed to all orders, as in a PSEG, to give reliable results.
Clearly it would be highly desirable to combine the two approaches.

It should be noted that for the first emission in a PSEG is typically
quite easy to modify the splitting functions to correctly reproduce
the exact matrix element, effectively replacing $c\sup{PS}_{11}$ with
$c\sup{ME}_{11}$, and in most PSEGs this is the default behavior for
most primary
sub-processes\cite{Gustafson:1988rq,Lonnblad:1992tz,Bengtsson:1987hr,
  Bengtsson:1987et,Seymour:1995we,Seymour:1995df,Seymour:1994ti,
  Lonnblad:1995wk,Lonnblad:1996ex,Miu:1998ju,Mrenna:1999mq,
  Corcella:1999gs}.

\subsection{The original CKKW procedure}
\label{sec:orig-ckkw-proc}

Comparing eqs.\ \ref{eq:psexpansion} and \ref{eq:meexpansion} the
solution should be obvious. Use a MEG to generate up to $N$ additional
partons above some cutoff, $y\sub{cut}$, but reweight the generated
states with the Sudakov form factors, and then add a parton shower
with the requirement that no partons above $y\sub{cut}$ are emitted.
This is the essence of the CKKW procedure. To calculate the Sudakov
form factors we need an ordered set of emission scales, which is not
provided by the MEG, since there all possible diagrams are added
coherently and emission scales are not well defined. In the original
CKKW procedure, the \kT clustering
algorithm\cite{kTalgorithm,Catani:1991hj} was used to define an
ordered set of scales which were used to analytically calculate
the Sudakov form factors. A reweighting was also done to have the
constructed scales as argument to \as. The resolution variable of
the \kT-algorithm was also used for the cutoff in the MEG, which
is not the same as the evolution variable in the PSEG. To ensure a
full coverage of the phase space the parton shower was therefore added
with the maximum scale as starting point, but vetoing all emissions
corresponding to the \kT-algorithm resolution variable above
$y\sub{cut}$ to avoid double-counting.

It was shown that this procedure removes the dependence on the MEG
cutoff, $y\sub{cut}$, to next-to-leading logarithmic accuracy.
However, there was still a clearly visible discontinuity in some
generated distributions. In \cite{Lonnblad:2001iq} the procedure was
improved in several ways when it was implemented for the \ariadne
program.

\subsection{The Dipole Cascade Model}
\label{sec:dipole-cascade-model}

\ariadne implements the dipole cascade model which is quite different
from conventional parton cascades. Rather than iterating $1\to2$ parton
splittings, gluons are emitted from colour-dipoles between
colour-connected partons resulting in $2\to3$ parton splittings. This
model has several advantages. Since gluons are emitted coherently by
colour-connected partons, there is no need for explicit angular
ordering. In addition, the evolution variable is defined as a
Lorentz-invariant transverse momentum which also is a suitable scale
to be used in \as. The evolution variable is defined as (for massless
partons)
\begin{equation}
  \label{eq:invpt}
  \pT^2=\frac{s_{12}s_{23}}{s_{123}},
\end{equation}
where parton 2 is the emitted one and $s_{ij}$ and $s_{ijk}$ are the
squared invariant masses of the two- and three-parton combinations.

The probability for a given emission is given in terms of the dipole
splitting functions, which depend on $\pT^2$ and a Lorentz invariant
rapidity defined as
\begin{equation}
  \label{eq:invy}
  y=\frac{1}{2}\ln\frac{s_{12}}{s_{23}}.
\end{equation}
The probability of a gluon emission from a dipole between two partons
$i,j$ is then given by
\begin{equation}
  \label{eq:probdip}
  dP(\pT^2,y)=\as(\pT^2)D_{ij}(\pT^2,y)
  \exp\left(-\int_{\pT^2}\frac{d\pT^{'2}}{\pT^{'2}}\int dy'
    \as(\pT^{'2})D_{ij}(\pT^{'2},y')\right)\frac{d\pT^2}{\pT^2}dy,
\end{equation}
where $\exp(\ldots)$ is the Sudakov form factor. The dipole splitting
functions, $D_{ij}$, depends on which partons are involved according
to
\begin{eqnarray}
  D_{\q\qbar}(\pT^2,y)&=&\frac{2}{3\pi}\frac{x_1^2+x_3^2}{(1-x_1)(1-x_3)}
  \label{eq:dipsplitqq}\\
  D_{\q\g}(\pT^2,y)&=&\frac{3}{4\pi}\frac{x_1^2+x_3^3}{(1-x_1)(1-x_3)}
  \label{eq:dipsplitqg}\\
  D_{\g\g}(\pT^2,y)&=&\frac{3}{4\pi}\frac{x_1^3+x_3^3}{(1-x_1)(1-x_3)}
  \label{eq:dipsplitgg}
\end{eqnarray}
where $x_i$ are the resulting energy fractions of the emitting partons
in the original dipole rest system, $x_i=2E_i/\sqrt{s_{123}}$, related to $\pT^2$ and $y$ according to
\begin{equation}
  \label{eq:ptyx1x3}
  y=\frac{1}{2}\ln\left(\frac{1-x_3}{1-x_1}\right),\qquad
  \pT^2=s_{123}(1-x_1)(1-x_3).
\end{equation}
It can be shown that the dipole splitting functions
are equivalent to the standard Altarelli--Parisi
splitting functions in the relevant soft and collinear
limits\cite{Gustafson:1988rq}. We also note that the $D_{\q\qbar}$
exactly corresponds to the leading order $\ee\to\q\g\qbar$ tree-level
matrix element.

Another feature of the dipole cascade, which will turn out to be
important for the CKKW implementation, is that all partons are always
on shell throughout the cascade. This is possible since the recoil
from the emitted parton can be absorbed by the two emitting ones. In
contrast, a conventional parton cascade does not have on-shell
intermediate states, and the full kinematics of an event is not
constructed until all the scales in the complete shower have been
generated.  While the energy loss of the emitting partons are defined
in the splitting functions in
\eqsref{eq:dipsplitqq}--(\ref{eq:dipsplitgg}), the transverse recoil
is chosen according to some principles detailed in
\cite{Lonnblad:1992tz}.

It can be noted that the ``inverse'' of the dipole cascade is a
well-behaved jet clustering algorithm. In fact such an algorithm has
been constructed, the \diclus algorithm\cite{Lonnblad:1993qd}, based
on successive clusterings of three jets into two, using the \pT in
\eqref{eq:invpt} as resolution scale, which has been shown to have
many attractive features\cite{Moretti:1998qx}.

There are, however, also some disadvantages with the dipole cascade
model, the main one being that it only deals with gluon emissions, and
the splitting of gluons into $\q\qbar$ pairs must be added by hand,
both for final-state\cite{Andersson:1990ki} and
initial-state\cite{Lonnblad:1995wk} splitting.

The initial-state splitting will be described in section
\ref{sec:sea-quark-emissions} below. Final-state splittings are simply
added as a possibility for a dipole connected to a gluon to split this
gluon into a \q\qbar-pair in addition to emit a gluon. Here, the
standard Altarelli--Parisi splitting function is used, divided between
the two dipoles connected to the gluon. This will result in a new
dipole splitting function,
\begin{equation}
  \label{eq:dipsplitQQ}
  D_{i\g}^{\q\qbar}=\frac{\xi}{4\pi}\frac{(1-x_2)^2+(1-x_3)^2}{1-x_1}.
\end{equation}
In the original formulation, the splitting was divided equal between
the two dipoles connected to the gluon, ie.\ $\xi=0.5$. However, in
the current \ariadne implementation, a larger fraction is given to the
smaller of the dipoles $i\g$ and $\g j$, hence for the $i\g$ dipole we
have $\xi=s_{\g j}/(s_{\g j}+s_{i\g})$.

Although these
splitting do not come in naturally in the dipole picture, they can be
incorporated in a consistent way and the resulting implementation in
\ariadne is probably the best model for describing both \tee final
states at LEP \cite{Hamacher:1995df}, and DIS final states at
HERA\cite{Brook:1995nn}.

\subsection{\protect\ariadne and CKKW}
\label{sec:ariadne-ckkw}

The \ariadne implementation of CKKW for \tee is described in detail in
\cite{Lonnblad:2001iq}. Considering the nature of the dipole cascade
it may seem reasonable to use the dipole clustering algorithm in
\diclus to construct scales, rather than using the \kT-algorithm.
However, in the \ariadne implementation a further step is taken. For
each partonic state generated by a MEG, all possible dipole cascade
histories are constructed, basically answering the question
\textit{how would \ariadne have generated this state?} A specific
history is then picked by weighting possible histories with the
product of the corresponding dipole splitting functions. The
implementation depends on the MEG generating specific colour
connections among the partons. Although this information is not
physical, it is usually provided by ME generator programs. (See
discussion in ref.~\cite{Boos:2001cv}). In principle one could
choose between all possible colour connections in the same way as
different histories are considered, but as most MEGs supply colour
information this is not necessary.

From a MEG generated state with $n$ additional partons, we can now
construct, not only an ordered set of emission scales,
$\pTi{n}^2,\pTi{n-1}^2,\ldots\pTi{1}^2$, but also the corresponding
set of intermediate states, $S_{n-1},S_{n-2},\ldots S_{0}$. As in the
standard CKKW, the reweighting with the correct scales in \as\ is done
using just the constructed scales.  However, the Sudakov form
factors are calculated using the fact that
$\sud{i}(\pTi{i}^2,\pTi{i+1}^2)$ exactly corresponds the probability
that no emission occurred from state $S_i$ between the scales
$\pTi{i}^2$ and $\pTi{i+1}^2$. Hence, letting \ariadne make a trial
emission, starting from the state $S_i$ with $\pTi{i}^2$ as the
maximum scale and throwing away the MEG event if the trial emission
was above $\pTi{i+1}^2$, will exactly correspond to reweighting with
the same Sudakov form factor, $\sud{i}(\pTi{i}^2,\pTi{i+1}^2)$, which
\ariadne would have used when generated the event.

A special treatment is given the trial emission from the MEG generated
state, $S_n$. Rather than using the cutoff, $\pTi{c}^2$, from the
dipole cascade, the event is thrown if the emission is above the
cutoff used in the MEG, while if the emission is below, the emission
is kept and the cascade is continued down to $\pTi{c}^2$ to produce a
full ME+PS event. In addition, if $n=N$, the highest number of
additional partons generated from the MEG, the emission from $S_N$ is
always kept. This was not done in the original CKKW prescription, but
is clearly needed, since otherwise we would never get events with
$N+1$ additional partons above the cutoff. In later developments of
CKKW such a treatment has been
added\cite{Krauss:2004bs,Mrenna:2003if,Schaelicke:2005nv}.

The whole procedure looks as follows:
\begin{enumerate}
\item First the number of partons, $n\le N$, to be generated is chosen
  according to the integrated tree-level matrix elements in the MEG,
  using a cutoff $y\sub{cut}$ in some jet resolution scale and a fixed
  $\as(\pTi{c}^2)$.
\item Then MEG is told to generate the momenta of the state with $n$
  additional partons according to the tree-level matrix element.
  Since we do not want any events below the cutoff in the dipole
  cascade, the invariant $\pT^2$ of the partons is checked, and if
  anyone is below $\pTi{c}^2$, the state is rejected and the procedure
  is restarted at step 1.
\item Now, all the intermediate states $S_{n-1}$, \ldots, $S_{0}$ and
  scales $\pTi{n}^2$, \ldots, $\pTi{1}^2$ are constructed
  corresponding to a possible dipole shower history of the generated
  $S_n$ state.
\item The generated event is rejected and we restart at step 1 with a
  probability given by $\prod_i\as(\pTi{i})/\as(\pTi{c})^n$ to get the
  correct scales in \as
\item We now make a trial emission with the dipole cascade from the
  state $S_0$ starting from the maximum scale, typically limited by
  the squared center-of-mass energy. If this emission is at a scale
  above $\pTi{1}^2$, the event is rejected and we restart from step 1.
  If not, a trial emission is performed from the state $S_1$ with a
  maximum scale of $\pTi{1}^2$.  If this emission is at a scale above
  $\pTi{2}^2$ the event is rejected and we restart from step 1. This
  procedure is repeated for all states down to $S_{n-1}$.  If no
  rejection has been made, a trial emission is made from the
  ME-generated state with $n$ additional partons starting from the
  scale $\pTi{n}^2$.  There are now two cases:
  \begin{anumerate}\itemsep 0mm
  \item If $n=N$ the trial emission is always kept and the dipole
    cascade is allowed to continue down to the cutoff $\pTi{c}^2$ and
    the event is accepted.
  \item If $n<N$, and all parton pairs pass the cut, $y\sub{cut}$ used
    in the MEG, the event is rejected and we restart from step 1. If
    any of the partons fail the cut, the trial emission is accepted
    and the dipole cascade is allowed to continue down to the cutoff
    $\pTi{c}^2$ and the event is accepted.
  \end{anumerate}
\end{enumerate}

Taken together, this so-called \emph{Sudakov veto} algorithm, giving
``exact \ariadne Sudakovs'', and the special treatment of the states
with highest multiplicity, results in a dramatic improvement of the
smoothness of observables sensitive to the cutoff region for \tee
events\cite{Lonnblad:2001iq}.

\subsection{CKKW in hadronic collisions}
\label{sec:ckkw-hadr-coll}

To extend this algorithm to be used for hadronic collisions is
conceptually straight forward. In \cite{Krauss:2004bs} and
\cite{Mrenna:2003if} the standard CKKW procedure was extended to
hadronic collisions in general and for the $\W+$jets process in
particular, and implemented for the \apacic, \herwig and \pythia
PSEGs. The principle is the same as for \tee. Jet construction is
done with the \kT-algorithm modified for hadronic collisions according
to \cite{Catani:1993hr}. The resulting ordered set of scales is
used in the analytic Sudakov form factors (in \cite{Mrenna:2003if} a
Sudakov veto algorithm similar to the one in \ariadne was used for the
\pythia implementation), and the events from the MEG was reweighted
with these and the properly scaled \as. The MEG states with highest
multiplicity was treated in the same way as in the \ariadne
implementation above.

There are some issues which need to be treated with special care. In a
PSEG, the initial states emissions are generated in a backward
evolution procedure which besides the standard partonic splitting
functions also involve ratios of PDFs. The leading order cross section
is given by
\begin{equation}
  \label{eq:sighatw}
  d\sigma_0=d\sigma_{hh\to \W}=
  \sum_{\q,\q'}xf_{\q}(x_+,\mu^2)xf_{\q'}(x_-,\mu^2)
  \hat{\sigma}_{\q\q'\to \W}(x_+x_-S)\frac{dx_+}{x_+}\frac{dx_-}{x_-},
\end{equation}
where the scale $\mu^2=m_\W^2=x_+x_-S$. Making one step in the backward
evolution with a $\g\to\q\qbar$ splitting will be performed with a
probability
\begin{equation}
  \label{eq:backwardstep}
  dP(Q^2,z)=\as P_{\g\to\q}(z)
  \frac{\frac{x_+}{z}f_{\g}(\frac{x_+}{z},Q^2)}{x_+f_{\q}(x_+,Q^2)}
  \sud{}(Q^2_{\max},Q^2)\frac{dQ^2}{Q^2}dz,
\end{equation}
where the Sudakov form factor can be formulated both with (\pythia)
and without (\herwig) ratios of PDFs. Both choices are formally
equivalent in the leading-log approximation, but only the former
choice corresponds exactly to a no-emission probability.  The maximum
scale is typically given by $m_\W^2$, and the Sudakov form factor
corresponds to a leading-log DGLAP resummation\footnote{Also
  next-to-leading logarithmic Sudakov form factors may be used
  \cite{Catani:2001cc}, although these may become larger than unity,
  disabling the interpretation as no-emission probabilities.}.
However, clearly there is nothing in the real world preventing
emissions with a scale above $m_\W^2$, and such parton states will be
generated by the MEG. In \cite{Krauss:2004bs} states with one or more
partons emitted at a scale above $m_\W^2$ are treated as coming from a
separate class of primary sub-processes, and the Sudakov form factors
are added only for additional emissions with a maximum scale given by
the smallest of the constructed scales above $m_\W^2$.

Although the resulting procedures in \cite{Krauss:2004bs} and
\cite{Mrenna:2003if} are shown to be fairly cutoff insensitive, there
is still some dependence, and it may be worth while to extend also the
\ariadne implementation of CKKW to hadronic collisions to see if a
better result can be achieved. As for the standard CKKW this extension
is in principle straight forward. However, there are some tricky
issues, mainly to do with the treatment of parton densities, and to
describe how we deal with these, we first have to describe how hadronic
collisions are implemented in \ariadne.

\section{The Dipole Cascade Model for Incoming Hadrons}
\label{sec:dipole-model-incom}

In contrast to conventional parton shower models, the dipole cascade
model does not separate between initial- and final-state gluon
radiation. Instead, gluons are always emitted from final-state dipoles
as in the \tee case. The cleanest situation is in the DIS
electro-production case, where the leading order process is
$\el\q\to\el\q$, ie.\ a quark is being kicked out of a hadron. A gluon
may then be emitted from the colour-dipole between the struck quark
and the hadron remnant, using the same dipole splitting function as in
the \tee case in \eqref{eq:dipsplitqq}. There is one major difference
though. In the \tee case, both the quark and anti-quark can be
considered point-like, but in the DIS case only the struck quark is
point-like (at least up to the resolution scale, $Q^2$, of the
exchanged virtual photon) while the hadron remnant is an extended
object with a size of roughly one fermi. Just as for the
electromagnetic case, radiation of wavelengths much smaller than the
size of the antenna is suppressed.

\subsection{Gluon emission from an extended source}
\label{sec:gluon-emission-from}

In \cite{Andersson:1989gp} it was argued that only a fraction of the
hadron remnant is effectively taking part in the emission. For a gluon
emission at the scale $\pT^2$ this fraction is given by
\begin{equation}
  \label{eq:suppress}
  a(\pT)=\left(\frac{\mu}{\pT}\right)^\alpha,
\end{equation}
where $\mu$ parameterizes the inverse size of the remnant and $\alpha$
reflects the dimensionality of the emitter. If only that fraction of
the remnant momentum is allowed to take part in the emissions, this
corresponds to a sharp cut in the allowed phase space for gluon
emission, limiting the transverse momentum mainly in the remnant
direction according to
\begin{equation}
  \label{eq:phasecut}
  \pT<\frac{W a(\pT)}{e^{+y}+a(\pT)e^{-y}}.
\end{equation}

Here we note another major difference as compared to conventional
initial-state parton showers. From \eqref{eq:phasecut} it is easy to
see that the maximum scale is given by
\begin{equation}
  \label{eq:ptmaxdis}
  \pTi{\max}=\left(\frac{W^2\mu^\alpha}{4}\right)^{\frac{1}{2+\alpha}},
\end{equation}
which may be much larger than $Q^2$ which is used as the maximum scale
in conventional PSEGs, especially for small $x$ values since
$W^2\approx Q^2/x$. Since the maximum scale is also used in the
Sudakov form factors, these do not only correspond to a standard DGLAP
resummation of leading logarithms of $Q^2$, but in the dipole model
they also resum, at least partially, logarithms of $1/x$. Note
however, that there is no formal equivalence to BFKL or CCFM
evolution. There is another similarity though. Even though the
emissions in the dipole model are ordered in \pT, they are not ordered
in rapidity, and conversely, following the emissions in rapidity from
the struck quark the transverse momenta of the gluons will be
unordered as in BFKL and CCFM.

In the implementation in \ariadne, the sharp cutoff in \pT is replaced
by a smooth function, $\Theta(\pT^2,y)$, with some power suppression
for emissions violating \eqref{eq:phasecut}. Concentrating on the soft
and collinear limits, where the splitting function is simply $\propto
d\ln\pT^2 dy$, we can write the probability of emitting a gluon as
\begin{equation}
  \label{eq:disgdip}
  dP(\pT^2,y)=\frac{4\as}{3\pi} \Theta(\pT^2,y)\sud{}(W^2,\pT^2)
  \frac{d\pT^2}{\pT^2}dy
\end{equation}
Comparing this to the corresponding initial-state $\q\to\q$ splitting
in a conventional parton shower, (cf.~\eqref{eq:backwardstep}) where
we have
\begin{equation}
  \label{eq:backwardstepqq}
  dP(Q^2,z)=\frac{4\as}{3\pi} \frac{1}{1-z}
  \frac{\frac{x_+}{z}f_{\q}(\frac{x_+}{z},Q^2)}{x_+f_{\q}(x_+,Q^2)}
  \sud{}(Q^2_{\max},Q^2)\frac{dQ^2}{Q^2}dz,
\end{equation}
and noting that in this limit
\begin{equation}
  \label{eq:zQ2jac}
  \frac{1}{z(1-z)}\frac{dQ^2}{Q^2}dz=\frac{d\pT^2}{\pT^2}dy,
\end{equation}
we see that the suppression function, $\Theta$, corresponds to the
ratio of PDFs
\begin{equation}
  \label{eq:suppratq}
  \Theta(\pT^2,y)\to
  z\frac{\frac{x_+}{z}f_{\q}(\frac{x_+}{z},Q^2)}{x_+f_{\q}(x_+,Q^2)}
\end{equation}

The fact that only a part of the remnant takes part in a gluon
radiation also means that only a fraction of it will obtain a
transverse recoil in an emission. This is handled by the addition of
so-called recoil gluons and is described in some detail in
\cite{Lonnblad:1992tz}. These recoil gluons will not be relevant for
this report, however they will play a role when implementing CKKW in
\ariadne for DIS and we will come back to them in more detail in a
future publication\cite{nilsprep}.

For W production in hadronic collisions, the primary sub-process is
$\q\qbar\to \W$ and the initial dipole from which gluons are radiated
is between the two remnants. The model is now the same as in DIS.
However, since both remnants are extended, the cutoff in
\eqref{eq:phasecut} becomes
\begin{equation}
  \label{eq:phasecutw}
  \pT<\frac{W a_1(\pT)a_2(\pT)}{a_2(\pT)e^{+y}+a_1(\pT)e^{-y}},
\end{equation}
and the maximum scale is
\begin{equation}
  \label{eq:ptmaxw}
  \pTi{\max}=\left(\frac{W^2\mu_1^{\alpha_1}\mu_2^{\alpha_2}}{4}\right)%
  ^{\frac{1}{2+\alpha_1+\alpha_2}}.
\end{equation}
Again the sharp cutoff is replaced by a power suppression of
transverse momenta above the limit in \eqref{eq:phasecutw}. Rather
than introducing recoil gluons to absorb the transverse recoil, we
note that any emission from the primary dipole corresponds to
initial-state radiation, for which it is natural that the recoil is
taken by the \W\ (otherwise it would not be possible to produce a \W\ 
with non-zero transverse momentum). In \cite{Lonnblad:1996ex} the
choice was to always transfer the transverse recoil to the \W\ in the
first emission, while in subsequent emissions, the recoil is only
transferred if the emitted gluon is close to the \W\ in phase space.

\subsection{Sea-quark emissions from remnant dipoles}
\label{sec:sea-quark-emissions}

Besides gluon radiation, there is also a possibility that one of the
quarks fusing into the \W\ is a sea-quark, in which case it could have
come from a perturbative splitting of a gluon as in
\eqref{eq:backwardstep}. As for the final-state gluon splitting into
\q\qbar, this process does not come in naturally in the dipole model.
Instead it is added by hand as an explicit initial-state splitting.
This procedure is detailed in \cite{Lonnblad:1995wk}, and is based on
different treatments of the remnants depending on whether a valence-
or a sea-quark entered into the primary sub-process. A sea-quark is
picked with the probability
$xf_{s\q}(x,Q^2)/(xf_{s\q}(x,Q^2)+xf_{v\q}(x,Q^2))$ and in this case
the complex remnant containing the anti-sea-quark and the valence
quarks is split into a colour-singlet hadron containing the
anti-sea-quark and a simple remnant. The sharing of longitudinal
momentum is inspired by the string fragmentation function as explained
in detail in \cite{Andersson:1989gp}. A dipole connected to such a
remnant is now allowed to emit the anti-sea-quark in a way similar to
a standard initial-state parton shower, except that the ordering is in
transverse momentum, changing \eqref{eq:backwardstep} to
\begin{equation}
  \label{eq:backwardsea}
  dP(\pT^2,z)=\as P_{\g\to\q}(z)
  \frac{\frac{x_+}{z}f_{\g}(\frac{x_+}{z},\pT^2)}{x_+f_{\q}(x_+,\pT^2)}
  \sud{}(\pTi{\max}^2,\pT^2)\frac{d\pT^2}{\pT^2}dz.
\end{equation}
As for the case of final-state gluon splitting, there is now in the
\W-production case several competing processes which can occur in the
primary dipole, and after generating one emission of each, the one
which gave the largest $\pT$ is chosen. If the emission of an
anti-sea-quark is chosen, it will form a new dipole with the remnant
of the hadron to which it previously belonged. The transverse recoil
is taken by hard subsystem, just as in a standard parton shower, where
the hard subsystem in our case is the \W\ and any other parton which
has been previously emitted.

It may seem counterintuitive that the essentially non-perturbative
splitting of the remnant is allowed to affect perturbative emissions.
However, the splitting will mainly influence the region very close to
the remnant, which is typically out of reach for current experiments.
Nevertheless, we will investigate alternative treatments in a future
publication.

Emitting an anti-sea-quark means that we have now extracted a gluon
from the incoming hadron. In a standard parton shower scenario it
would then be possible to evolve this gluon backwards either with a
$\g\to\g$ splitting or a $\q\to\g$ splitting. In \ariadne, the former
is modeled by gluon emissions from either of the two dipoles connected
to the two remnants. In this case we use the same suppression function
as in \eqref{eq:disgdip}, but comparing to \eqref{eq:backwardsea} we
find that it now corresponds to the ratio of gluon densities,
\begin{equation}
  \label{eq:suppratg}
  \Theta(\pT^2,y)\to
  \frac{\frac{x_+}{z}f_{\g}(\frac{x_+}{z},Q^2)}{x_+f_{\g}(x_+,Q^2)},
\end{equation}
where the extra factor $z$ in \eqref{eq:suppratq} is absent since this
is now included in the gluon splitting function. The initial-state
$\q\to\g$ splitting is not included in the \ariadne program, but it
could in principle be added in the same way as the sea-quark emission.
Again we will investigate this in a future publication.

Clearly, the dipole model for incoming hadrons has some conceptual
problems, especially when it comes to initial-state $\g\to\q$
splittings. However, it also has some advantages. The first emission
is quite easily modified to correctly reproduce the leading order
matrix element, both for DIS and \W-production. Also, a larger part of
phase space is available for gluon emissions as compared to DGLAP
based initial-state parton showers. This enables \ariadne to reproduce
small-$x$ observables in DIS, such as the forward jet rates, where no
conventional shower succeeds. Below we shall also see that \ariadne
gives a somewhat harder peak in the W \pT-spectrum than conventional
parton showers, which we know peaks below the data measured at the
Tevatron.

Now that we have explained how \ariadne handles \W-production in
hadronic collisions, we can proceed with describing how to combine it
with a fixed-order tree-level MEG. As for the \tee case it will
involve constructing all possible cascade histories of a produced
MEG state, the reweighting with Sudakov form factors using a
Sudakov-veto algorithm, and finally the reweighting with \as\ as well
as with ratios of PDFs and the suppression function,~$\Theta$.

\section{\protect\ariadne and CKKW for \boldmath\W\ production}
\label{sec:prot-ckkw-boldm}

The states delivered by a MEG contains information about the momenta,
colour connections and types of the incoming and outgoing particles in
the generated sub-process. The states are generated according to exact
tree-level matrix elements, using a fixed \as\ evaluated at some scale
$Q_0^2$ and weighted by the relevant parton densities typically
evaluated at the same scale. $Q_0^2$ is usually taken to be the cutoff
scale used to regularize soft and collinear divergencies.

\subsection{Constructing the Emissions}
\label{sec:reconstr-emiss}

To construct a dipole cascade history of this state it is first
necessary to introduce the remnants so that all outgoing coloured
particles from the sub-process are connected with dipoles. This is
done in the same way as for standard \ariadne. In figure
\ref{fig:remnants} the different possible connections of dipoles to
the remnants are described schematically for \W-production in
$\p\pbar$ collisions. We see that if a gluon has been extracted from a
baryon (lower part of figure \ref{fig:remnants}b), there are two
remnants one containing a quark connected to parton in the end of
anti-colour line of the gluon, and one di-quark connected to the end
of the colour line. Furthermore, if a sea-quark is extracted from a
proton (upper part of figure \ref{fig:remnants}a), the parton in the
end of its colour line will be connected with a di-quark remnant,
while the anti-sea-quark will form a hadron together with the remaining
valence flavour, as described in section \ref{sec:sea-quark-emissions}
above.  Similarly, if an anti-sea-quark is extracted (upper part of
figure \ref{fig:remnants}b), the parton in the end of its colour line
will be connected with a single quark remnant, while the sea-quark
will form a hadron together with the remaining valence flavours.
Finally if a valence quark is extracted (lower part of figure
\ref{fig:remnants}a), the remnant is a di-quark which is connected
with the parton on the end of the colour line.

\FIGURE[t]{
  \begin{picture}(200,150)(0,0)
    \Line(20,10)(150,10)
    \Line(20,17)(150,17)
    \Line(20,24)(50,24)
    \GOval(20,17)(12,3)(0){0}
    \Oval(150,13.5)(7,3)(0)

    \Line(20,140)(150,140)
    \Line(20,133)(150,133)
    \Line(20,126)(150,126)
    \Line(50,119)(150,119)
    \GOval(20,133)(12,3)(0){0}
    \Gluon(45,126)(50,119){2.5}{1}
    \Oval(150,137)(6,3)(0)
    \GOval(150,123)(6,3)(0){0}

    \Line(50,119)(100,75)
    \Line(50,24)(100,75)
    \Vertex(100,75){1.5}

    \Photon(100,75)(150,75){2.5}{8}
    \DashCArc(100,75)(83,-48,48){2}

    \Text(150,110)[]{\scriptsize $[\qbar_{s}\q_{v1}]$}
    \Text(150,150)[]{\scriptsize $(\q_{v2}\q_{v3})$}
    \Text(150,65)[]{\scriptsize $\Wp$}
    \Text(150,30)[]{\scriptsize $(\qbar_{v2}\qbar_{v3})$}
    \Text(65,50)[]{\scriptsize $\qbar_{v1}$}
    \Text(65,95)[]{\scriptsize $\q_{s}$}
    \Text(10,17)[]{\scriptsize $\pbar$}
    \Text(10,133)[]{\scriptsize $\p$}
    \Text(75,160)[]{(a)}

  \end{picture}
  \begin{picture}(200,150)(0,0)
    \Text(75,160)[]{(b)}
    \Text(150,110)[]{\scriptsize $[\q_{s}\q_{v1}\q_{v2}]$}
    \Text(150,150)[]{\scriptsize $(\q_{v3})$}
    \Text(150,85)[]{\scriptsize $\Wp$}
    \Text(10,17)[]{\scriptsize $\pbar$}
    \Text(10,133)[]{\scriptsize $\p$}
    \Text(140,35)[]{\scriptsize $(\qbar_{v2}\qbar_{v3})$}
    \Text(150,0)[]{\scriptsize $(\qbar_{v1})$}
    \Text(65,95)[]{\scriptsize $\qbar_{s}$}
    \Text(75,65)[]{\scriptsize $\q$}
    \Text(150,55)[]{\scriptsize $\qbar$}
    \Line(20,10)(150,10)
    \Line(20,17)(150,17)
    \Line(20,24)(150,24)
    \GOval(20,17)(12,3)(0){0}
    \Oval(150,21)(6,3)(0)
    \Oval(150,10)(3,3)(0)

    \Line(20,140)(150,140)
    \Line(20,133)(150,133)
    \Line(20,126)(150,126)
    \Line(50,119)(150,119)
    \GOval(20,133)(12,3)(0){0}
    \Gluon(45,126)(50,119){2.5}{1}
    \Oval(150,140)(4,3)(0)
    \GOval(150,126)(8,3)(0){0}

    \Line(50,119)(100,75)
    \Gluon(50,24)(70,50){2.5}{4}
    \Vertex(70,50){1.5}
    \Line(70,50)(100,75)
    \Line(70,50)(150,50)
    \Vertex(100,75){1.5}

    \Photon(100,75)(150,75){2.5}{8}
    \DashCArc(100,75)(85,-50,50){2}
    \DashCArc(141,36)(21,-48,48){2}
  \end{picture}
  \caption{\label{fig:remnants}Different ways dipoles are connected
    (dashed arcs) depending on which kind of parton is extracted from
    a baryon.  Filled ovals corresponds to colour-singlet hadrons,
    while open ovals represents coloured remnants.}}

The construction will now proceed iteratively, each step
corresponding to the inverse of an emission in the dipole cascade. All
possible constructions will be made, and afterwards one of them will
be picked. In each of the construction steps we must determine
\begin{itemize}
\item The scale of the corresponding emission.
\item The value of splitting function to be used to give different
  weights to different possible construction paths.
\item The ratio of PDFs or the value of the suppression function which
  would have been used in the corresponding emission. This will be
  used to reweight the events.
\item The way the momentum of the emitted parton is distributed among
  the emitters. There will usually be three partons constructed into
  two, which means that the total energy and momentum is always
  conserved with all partons staying on-shell. However, the
  orientation of the final two partons in the rest system of the
  construction needs to be specified.
\end{itemize}

Some of the possible construction steps correspond to emissions from
dipoles between partons from the hard sub-process, and these are the
same as in \tee.  Then there is a group of construction steps which
involve hadron remnants, which are particular to hadronic collisions
in general and to \W-production in particular. In appendix
\ref{sec:appendix} we present a complete list of all possible
construction steps

After the construction procedure we are normally left with several
possible cascade histories. Most of these will end up in a zeroth
order state containing only remnants and one \W\ with no transverse
momentum. There will be some diagrams generated by the MEG which never
could have come from an \ariadne cascade. One example is the
initial-state $\q\to\g$ splitting discussed in the end of section
\ref{sec:sea-quark-emissions}, in which case the constructed state
is accepted anyway. However, there are also diagrams, such as the one
in figure \ref{fig:impossible}, which could not be produced even by a
conventional parton shower. In this case the construction is stopped
before reaching the zeroth order state, and this state is then treated
as a separate leading order process and the reweighting is only
applied to additional partons (similarly to the treatment in
\cite{Krauss:2004bs} mentioned above).

\FIGURE[t]{
  \begin{picture}(220,130)(0,0)

    \Text(10,15)[]{\scriptsize \pbar}
    \Text(10,115)[]{\scriptsize \p}
    \Text(75,50)[]{\scriptsize \ubar}
    \Text(75,80)[]{\scriptsize \qu}
    \Text(155,75)[]{\scriptsize \dbar}
    \Text(205,90)[]{\scriptsize \ubar}
    \Text(205,65)[]{\scriptsize \Wp}
    \Text(205,40)[]{\scriptsize \qd}

    \Line(20,10)(200,10)
    \Line(20,15)(200,15)
    \Line(20,20)(60,20)
    \GOval(20,15)(12,3)(0){0}
    \Oval(200,12.5)(7,3)(0)

    \Line(20,120)(200,120)
    \Line(20,115)(200,115)
    \Line(20,110)(60,110)
    \GOval(20,115)(12,3)(0){0}
    \Oval(200,117.5)(7,3)(0)

    \Line(60,110)(100,65)
    \Line(60,20)(100,65)
    \Vertex(100,65){1.5}

    \Gluon(100,65)(140,65){2.5}{7}
    \Vertex(140,65){1.5}
    \Line(140,65)(200,40)

    \Line(140,65)(170,72)
    \Vertex(170,72){1.5}
    \Line(170,72)(200,90)
    \Photon(170,72)(200,72){2.5}{9}
  \end{picture}
  \caption{\label{fig:impossible}An example of a \W-strahlung diagram.
    Such diagrams are not modeled by standard \protect\ariadne.}}

The resulting alternative cascade histories may or may not have an
ordered set of constructed scales. When choosing a history according
to their weights given by the products of the splitting functions, we
first only consider true \ariadne histories with ordered scales. Only
if no such histories were found, the other histories are considered.
Histories corresponding to figure \ref{fig:impossible} will only be
considered if no full constructions are found.

\subsection{Reweighting the Events}
\label{sec:reweighting-events}

For a given MEG state, $S_n$, with $n$ additional jets, we have now
constructed a dipole cascade history with complete intermediate
states, $S_n,\ldots,S_0$, and the corresponding emission scales,
$\pTi{n},\ldots,\pTi{1}$, and we can proceed with the reweighting.

First we note that the MEG has used PDFs typically evaluated at the
cutoff scale, $Q_0^2$ with $x_+$ and $x_-$ given by the light-cone
momentum fractions of the partons, $i$ and $j$, entering the hard
sub-process. This should be compared with the starting point for a
normal parton cascade generation, where we just have a $\q\qbar'\to\W$
sub-process, and the PDFs are evaluated at the scale $m_\W^2$ and
$x_+'$ and $x_-'$ given by the corresponding momentum fraction for the
$\q$ and $\qbar'$. Our strategy is to follow the \ariadne cascade as
closely as possible, just replacing the product of dipole splitting
function with the exact tree-level matrix element, so to get the same
starting point, we take the $\q$ and $\qbar'$ of the
state $S_0$ and their $x_+'$ and $x_-'$, and reweight
the event with
\begin{equation}
  \label{eq:WPDF0}
  \omega_0 = \frac{x_+'f_\q(x_+',m_\W^2)\cdot x_-'f_{\qbar'}(x_-',m_\W^2)}
  {x_+f_i(x_+,Q_0^2)\cdot x_-f_j(x_-,Q_0^2)}.
\end{equation}
If a construction instead ended in a state such as the one in figure
\ref{fig:impossible}, the corresponding incoming partons and their
momentum fractions are used instead with the scale given by $m_H^2$,
the squared invariant mass of the hard sub-process.

Then we reweight with all the PDF ratios, $R\sup{PDF}_i$, determined
in the construction,
\begin{equation}
  \label{eq:WPDFi}
  \omega_1 = \prod_{i=1}^nR\sup{PDF}_i.
\end{equation}
This comes about since the exact tree-level ME used corresponds to the
product of splitting functions, while in a parton cascade we also have
ratios of PDFs as in eqs.~(\ref{eq:backwardstepqq}) and
(\ref{eq:backwardsea}). Depending on the emission, these ratios can be
either 1 for a final-state emission, the ratio of PDFs for the case of
initial-state $\q\to\g$ and $\g\to\q$ splittings, the suppression
function $\Theta$ for an initial-state $\g\to\g$ splitting and
$\Theta/z$ for an initial-state $\q\to\q$ splitting (cf.\ eqs.\ 
(\ref{eq:suppratq}) and (\ref{eq:suppratg})). We note that for a
conventional parton cascade, where the $\Theta$ functions would be
replaced by ratios of PDFs, the $\omega_0$ and $\omega_1$ weights
would basically cancel each other, which is why these did not show up
in the procedures in \cite{Krauss:2004bs} and \cite{Mrenna:2003if}.

We then reweight with the correct scales in \as\ according to
\begin{equation}
  \label{eq:Was}
  \omega_2=\frac{\prod_{i=1}^n\as(\pTi{i}^2)}{\as(Q_0^2)^n}.
\end{equation}
Again, for the situation in figure \ref{fig:impossible}, the first two
scales are taken to be $m_H^2$.

Finally we need to reweight with the Sudakov form factors in \ariadne.
This is done with the same Sudakov-veto algorithm as was presented in
section \ref{sec:ariadne-ckkw}. There are, however a few details which
should be mentioned.

The starting scale for the trial emission from the leading order
state, $S_0$, is given by $\pTi{\max}^2=W^2/4$, where $W$ is the total
invariant mass of the hadronic collision, ie.\ the same as for the
standard \ariadne treatment of \W\ production. For the situation in
figure \ref{fig:impossible}, $m_H^2$ is used instead.

If the constructed cascade history contains unordered scales, such
that $\pTi{i}^2<\pTi{i+1}^2$, the two corresponding emissions will be
treated as a combined emission with $\pTi{i+1}^2$ as the scale. The
Sudakovs will be generated with a trial emission from the state
$S_{i-1}$ with a minimum scale of $\pTi{i+1}^2$ and a trial emission
from the state $S_{i+1}$ with a maximum scale of $\pTi{i+1}^2$, while
there is no Sudakov generated from the state $S_i$.

Finally, in the trial emission from the state $S_n$, for $n<N$, when
checking if the resulting partons are above the jet cutoff used in the
MEG, possible recoil gluons are not considered. Such recoil gluons may
appear in \ariadne, but they are typically rather soft, and including
them would very often result in the emission being below the cutoff,
even if the emitted gluon is not.

\subsection{The Full Algorithm}
\label{sec:full-algorithm}

We can now summarize the full algorithm. The way it is used below
results in weighted events. This is because of the complicated
reweightings which takes place. However, all weights are positive, and
by carefully choosing the PDFs and \as\ used in the MEG, it should be
possible to have a vetoing procedure so that all events end up with
unit weight.

\begin{enumerate}
\item First the number of partons, $n\le N$, to be generated is chosen
  according to the integrated tree-level matrix elements in the MEG,
  using a cutoff $Q^2_0$ in the jet resolution scale given by the
  longitudinally invariant \kT-algorithm. A fixed \as\ is used and the
  PDFs are typically sampled at the $Q_0^2$ scale.
\item Then the MEG is told to generate the momenta of the state with
  $n$ additional partons according to the tree-level matrix element.
  Since we do not want any events below the cutoff in the dipole
  cascade, the invariant $\pT^2$ of the partons is checked, and if
  anyone is below $\pTi{c}^2$, the state is rejected and the procedure
  is restarted at step 1.
\item Now, all the intermediate states $S_{n-1}$, \ldots, $S_{0}$ and
  scales $\pTi{n}^2$, \ldots, $\pTi{1}^2$ are constructed according
  to the procedure in section \ref{sec:reconstr-emiss}, resulting in a
  possible dipole shower history of the generated $S_n$ state.
\item The event is reweighted by the weight factors given in
  \eqsref{eq:WPDF0}--(\ref{eq:Was}).
\item We now make a trial emission with the dipole cascade from the
  state $S_0$, starting from the maximum scale $\pTi{\max}^2=W^2/4$. If
  this emission is at a scale above $\pTi{1}^2$, the event is rejected
  and we restart from step 1.  If not, a trial emission is performed
  from the state $S_1$ with a maximum scale of $\pTi{1}^2$.  If this
  emission is at a scale above $\pTi{2}^2$ the event is rejected and
  we restart from step 1. This procedure is repeated for all states
  down to $S_{n-1}$.  If no rejection has been made, a trial emission
  is made from the ME-generated state with $n$ additional partons
  starting from the scale $\pTi{n}^2$.  There are now two cases
  \begin{anumerate}\itemsep 0mm
  \item If $n=N$ the trial emission is always kept and the dipole
    cascade is allowed to continue down to the cutoff $\pTi{c}^2$ and
    the event is accepted.
  \item If $n<N$, and all parton pairs pass the cut, $Q_0^2$, used in
    the MEG, the event is rejected and we restart from step 1. If any
    of the partons fail the cut, the trial emission is accepted and
    the dipole cascade is allowed to continue down to the cutoff
    $\pTi{c}^2$ and the event is accepted.
  \end{anumerate}
\end{enumerate}

\section{Results}
\label{sec:results}

To test our algorithm, we have generated \Wp+$n$jet events, with $n\le
N=4$ with the \madgraph/\madevent program\cite{Maltoni:2002qb} for a
\p\pbar\ collider at a total energy of 1960~GeV, ie.\ corresponding to
the Tevatron run II. The longitudinally invariant \kT-algorithm was
used\footnote{Using \texttt{MODE=4211} in the \ktclus
  program\cite{Catani:1993hr}.} to regularize the cross section, using
cutoffs $E\sub{cut}=12, 17$ and $22$~GeV. We used the
CTEQ6L\cite{Pumplin:2002vw} PDF parameterization using $E\sub{cut}^2$
as scale.  $E\sub{cut}^2$ was also used as the scale in \as. The event
was generated with unit weight and was then reweighted according to
the algorithm in section \ref{sec:full-algorithm}. To avoid wildly
fluctuating weights due to the ratios of PDFs, arising from situations
in which the events in \madgraph had large $x$ values where the PDF is
very small, the \Wp\ was required to have limited rapidity, $|y|<2.5$,
and all partons were required to have a limited
pseudorapidity, $|\eta|<2.5$. For the same reason, when constructing
the remnants, both the valence- and sea-quark alternatives were used
with appropriate weights, rather than choosing between the two as
described in section \ref{sec:sea-quark-emissions}.

In the following we will use the notation ME$N$PS for results from our
new algorithm using \Wp+0,\ldots,\Wp+$N$ jets from \madgraph (the
individual contributions from $\Wp+n$ jets are denoted ME$^n\!N$PS),
while \ariadne will denote results obtained with the default \ariadne
treatment.

As mention above, \ariadne by default already has a matrix element
correction for the first emission in \W\ production\footnote{In fact
  we discovered a small bug in the default \protect\ariadne treatment,
  related to a mismatch between the invariant \pT and the actual
  transverse momentum in gluon emissions.  This bug has been fixed in
  the latest release of \ariadne.}. Hence, as a first test of our new
algorithm is to run it with $N=1$, in which case we should get the
same result as the standard default \ariadne. In fact, had the
construction procedure been exact, the results would be exactly the
same. Of course, the construction can never be really exact, but for
only one additional jet it is fairly close.

\FIGURE[t]{
      \epsfig{file=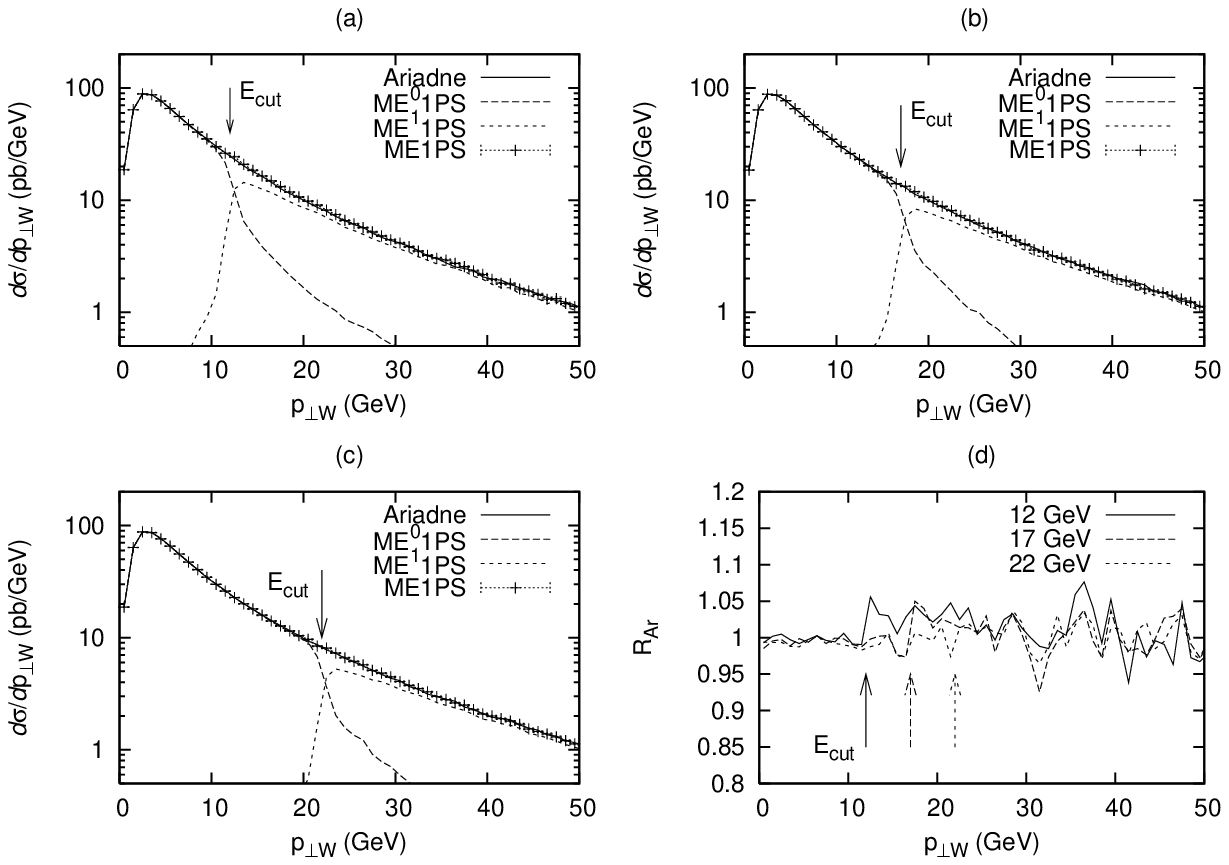,width=15cm}
  \caption{\label{fig:ptw01}
    Cross section as a function of transverse momentum of the W for 
    standard \protect\ariadne compared to the first order matrix element
    correction for 
    the different cutoffs (a) 12~GeV, (b) 17~GeV and (c) 22~GeV.
    Plot (d) shows the ratio between the distributions including 
    matrix element corrections and standard \protect\ariadne.}}

In figure \ref{fig:ptw01} we show the \pT\ spectrum of the \Wp\ for
the new algorithm with $N=1$, ME1PS, compared to default \ariadne.
Clearly the agreement is very good. In particular we note that there
is no significant discontinuity or other strange behaviors around the
different cutoffs used. This agreement is not trivial, since the ME1PS
curve is a sum of \Wp+0 and \Wp+1 jet event from \madgraph.  The fact
that there is a small contribution of \Wp+0 event above the cutoff
(and vice versa, some \Wp+1 events below the cutoff) is an effect of
the added cascade.  The effect of the cutoffs is not completely
invisible, however, as is clear in figure \ref{fig:ptw01}d, where we
enhance the effect by showing the ratio of \pT-spectra of ME1PS and
\ariadne for the different cutoffs.

We can now proceed with some confidence to investigate our new
algorithm also for higher parton multiplicities in the MEG. Here we
will, of course, expect differences w.r.t.\ \ariadne. These
differences would be the ones desired from replacement of the products
of splitting functions with the exact tree-level matrix elements, and
also from the additional processes not present in the dipole cascade,
such as the one in figure \ref{fig:impossible}. For small scales,
these differences should be small and we would still like to have a
smooth behavior of any observable sensitive to the cutoff used in
\madgraph, at least for small enough $E\sub{cut}$. There may also be
differences arising from deficiencies in our algorithm, since there are
now additional construction steps possible.

\FIGURE[t]{
      \epsfig{file=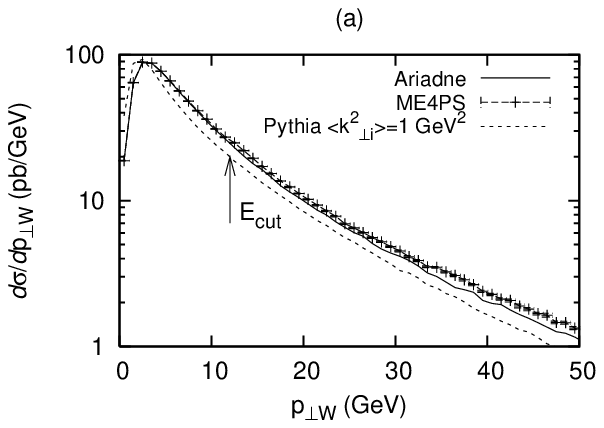,width=7.8cm}%
      \epsfig{file=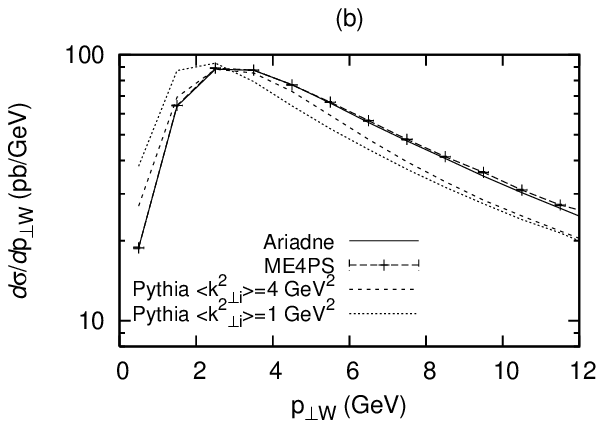,width=7.8cm}
  \caption{\label{fig:ptw04}
    Cross section as a function of transverse W momentum for standard
    \protect\ariadne, \protect\ariadne with fourth order matrix 
    element corrections and \protect\pythia with 1~GeV$^2$ and 
    4~GeV$^2$ intrinsic \kT}}

In figure \ref{fig:ptw04}a we again show the \Wp\ \pT\ spectrum, but
now using ME4PS and comparing with \ariadne and also with the default
\pythia parton shower (which also includes a tree-level matrix element
correction for the first emission). We find that there is now an
increase at large \pT\ for the ME4PS case, which is attributed to the
desired higher-order effects. We note that there is still no dramatic
discontinuity around the cutoff.

There is distinct difference in the small-\pT\ behavior for the
\pythia distribution. The peak in \pythia is shifted towards smaller
\pT, as compared to \ariadne, with or without the new matching
algorithm. This is a known problem with \pythia, which need an
uncomfortably large intrinsic transverse momentum of the proton to
reproduce data.  This is shown in more detail in figure
\ref{fig:ptw04}b where we focus on the small-\pT\ part of the spectrum
and where we have two curves for \pythia, one with an average squared
intrinsic transverse momentum, $\langle\kTi{i}^2\rangle$, of 1~GeV$^2$
and one with 4~GeV$^2$. Note that also \ariadne has an intrinsic
transverse momentum, but this is at a typical non-perturbative value
of $\langle\kTi{i}^2\rangle=0.36$~GeV$^2$, but the increased
possibility of radiating gluons in the dipole cascade, especially in
the direction of the remnants, will give the slightly harder \pT\ 
spectrum. From reference \cite{Thome:2004sk} we know that \pythia can
only describe data with the higher intrinsic transverse
momentum\footnote{In later \protect\pythia releases the higher value
  is the default.}, which is quite close to \ariadne\footnote{We have
  not compared directly with data here due to uncertainties about the
  corrections made to the data.} There is still a difference in shape
and with the increased statistics collected in Tevatron Run II, it may
be possible to distinguish between the two.

\FIGURE[t]{
      \epsfig{file=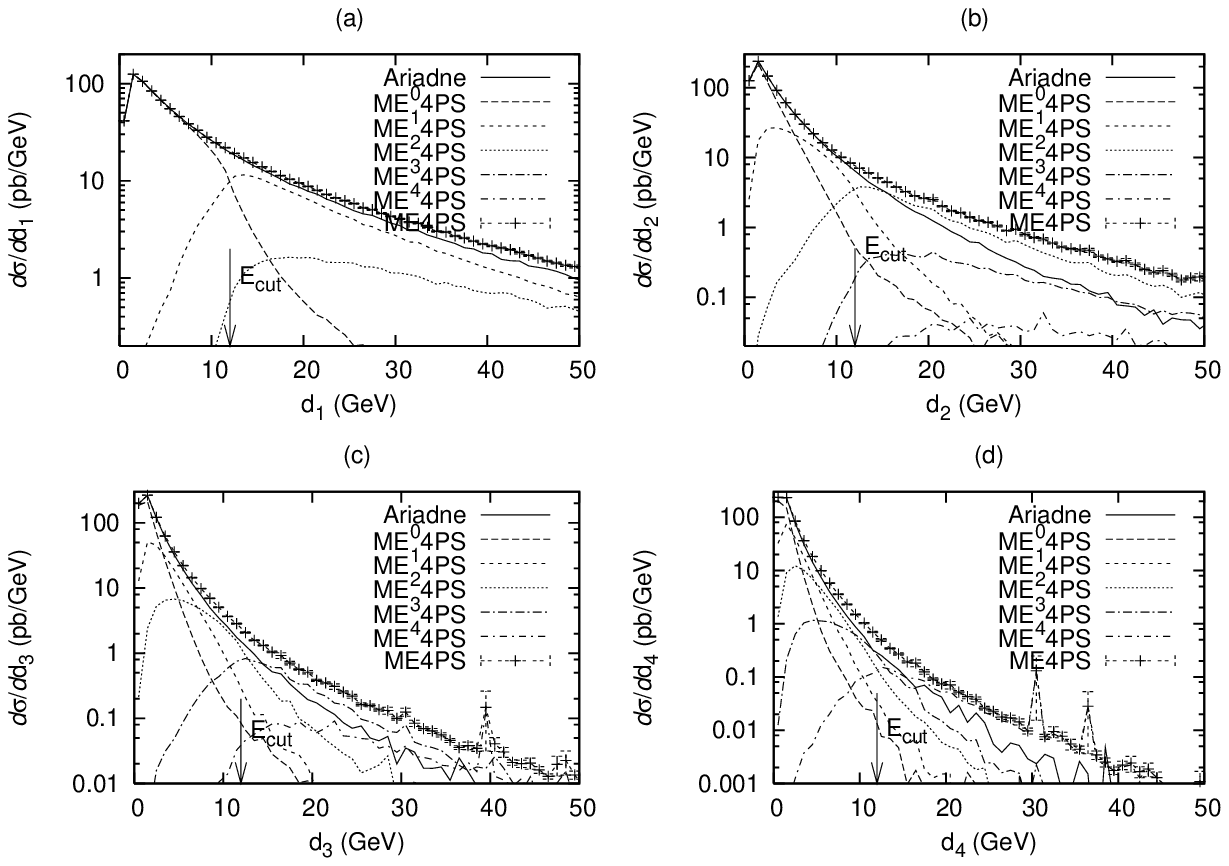,width=15cm}
  \caption{\label{fig:dn}
    Cross section as a function of the scale where the jets merge in
    the \kT-algorithm, where $d_i$ denotes the scale when $i$ jets
    merge to $i-1$ jets. Standard \protect\ariadne is compared to
    fourth order matrix element corrected distributions at a 12~GeV
    cutoff}}

\FIGURE[t]{
      \epsfig{file=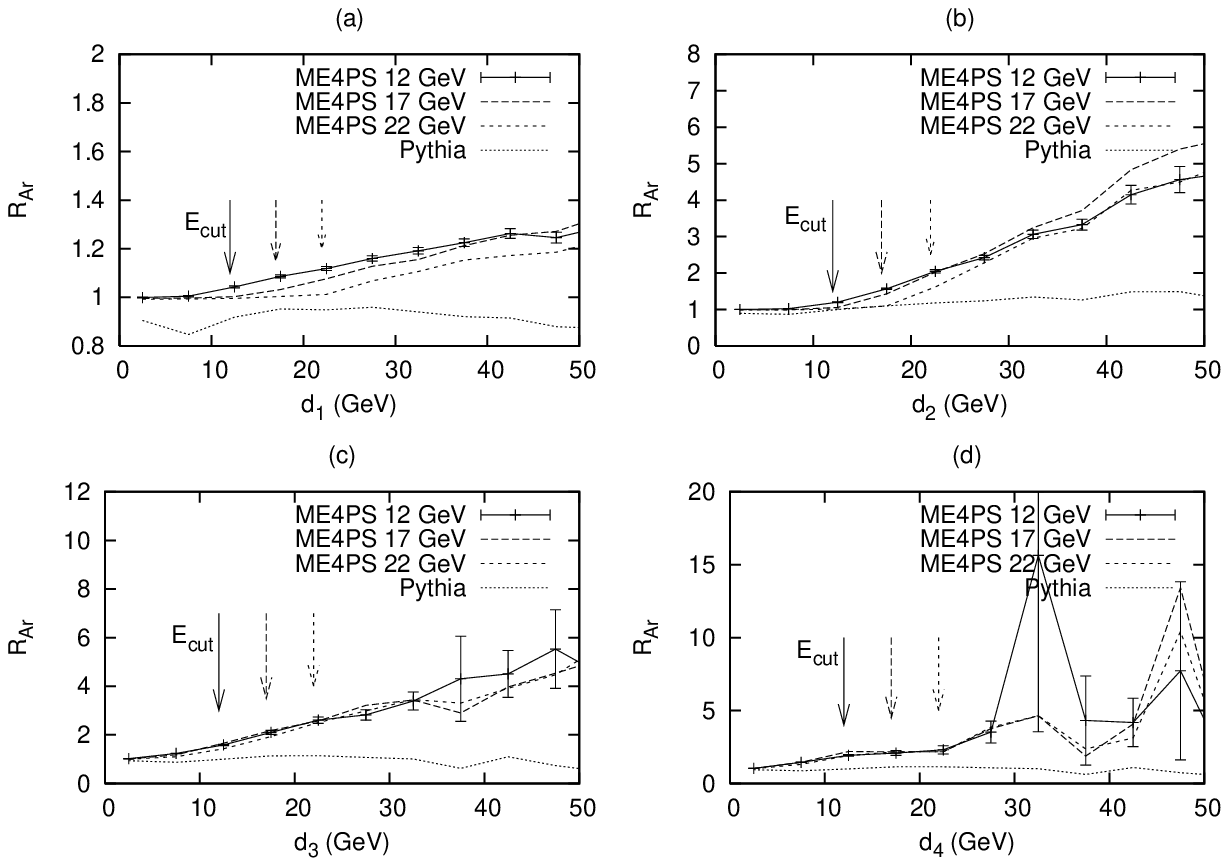,width=15cm}
  \caption{\label{fig:dnrat}
    ME corrected distributions and \protect\pythia divided by
    \protect\ariadne as a function of the jet merging scale in the
    \kT-algorithm for all different cutoffs used.}}

Next we want to check the cutoff sensitivity also for higher jet
rates. We do that by taking the final events on parton level and
cluster them with the same \kT-algorithm which was used for the
regularization in \madgraph, and then look at what value of the
resolution variable, $d_n$, an event is clustered from $n$-jets to
$n-1$-jets. In figure \ref{fig:dn} we show such distributions for
$n=1,2,3$ and 4 for $E\sub{cut}=12$~GeV. We also show the individual
contributions from different parton multiplicities delivered by
\madgraph. We see that there is a clear difference between ME4PS and
\ariadne for large $d_n$ values, which is expected from the improved
treatment of events with several hard jets. We note that there is a
rather smooth transition across $E\sub{cut}$. In figure
\ref{fig:dnrat} we also show results for $E\sub{cut}=17$ and 22~GeV,
now presented as ratios between ME4PS and \ariadne. As expected we
here see more clearly where the new matrix element treatment sets in
above $E\sub{cut}$. In figure \ref{fig:dnrat} we also show the ratio
between \pythia and \ariadne and we find that, as compared to the
effects of the matrix element corrections, the difference between the
two cascades is small.

\FIGURE[t]{
  \begin{picture}(200,150)(0,0)
    \Text(75,150)[]{(a)}
    \Text(150,105)[]{\scriptsize $\Wp$}
    \Text(10,17)[]{\scriptsize $\pbar$}
    \Text(10,133)[]{\scriptsize $\p$}
    \Line(20,10)(150,10)
    \Line(20,17)(150,17)
    \Line(20,24)(150,24)
    \GOval(20,17)(12,3)(0){0}
    \Oval(150,21)(6,3)(0)
    \Oval(150,10)(3,3)(0)

    \Line(20,140)(150,140)
    \Line(20,133)(150,133)
    \Line(20,126)(50,126)
    \Line(50,119)(50,119)
    \GOval(20,133)(12,3)(0){0}
    \Oval(150,136.5)(6,3)(0)

    \Line(50,126)(80,95)
    \Gluon(50,24)(70,70){2.5}{4}
    \Vertex(70,70){1.5}
    \Line(70,70)(80,95)
    \Line(70,70)(100,70)
    \Vertex(80,95){1.5}
    \Vertex(59,41){1.5}
    \Gluon(59,41)(150,41){2.5}{10}
    \Vertex(62,55){1.5}
    \Gluon(62,55)(100,55){2.5}{4}
    \Vertex(75,82.5){1.5}
    \Gluon(75,82.5)(100,82.5){2.5}{3}

    \Photon(80,95)(150,95){2.5}{8}
    \Text(150,68)[]{\scriptsize $\Delta\eta_{\W j}$}
    \LongArrow(150,72)(150,93)
    \LongArrow(150,64)(150,43)
  \end{picture}\epsfig{file=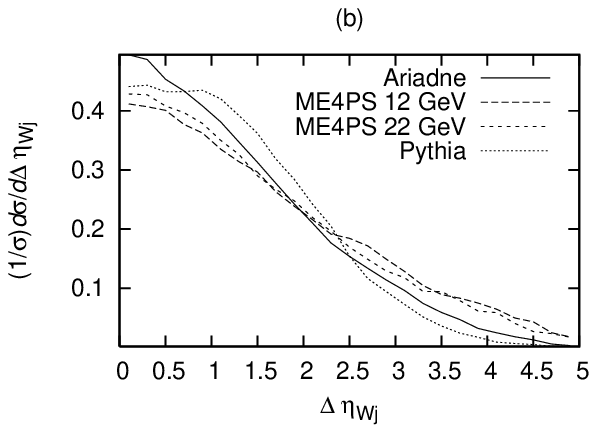,width=8cm}
  \caption{\label{fig:etaWj}
    Part (a) shows a typical diagram contributing to the
    pseudorapidity difference between the W and the hardest jet.
    Diagrams with soft jets between the W and the hardest jet can not
    be generated through DGLAP evolution, but are included in the
    matrix element corrections.  Plot (b) is a normalized distribution
    of the difference in pseudorapidity between the hardest jet and
    the W, where the jet is defined using the \kT-algorithm with a
    12~GeV cutoff and the hardest jet has a transverse momentum
    greater than 40~GeV.  The distribution is shown for
    \protect\ariadne (full line), \protect\pythia (dotted line), and
    ME4PS with a 12~GeV and 22~GeV cutoff (long- and short-dashed
    lines respectively).}}

Next we want to see if the features of the \ariadne resummation are
reflected in our new algorithm. As noted before, the emission of
gluons are allowed in a larger phase space region in \ariadne as
compared to a conventional PSEG, hence the no-emission probabilities
should be affected. Also, in a conventional DGLAP-based initial-state
PSEG, the parton closest to the W is also the hardest one. This is not
the case for \ariadne where, effectively, contributions of emissions
with lower \pT\ between the hardest parton and the W is taken into
account, as illustrated in figure \ref{fig:etaWj}a. One observable
which may be sensitive to this difference is the pseudorapidity
difference between the \W\ and a jet, $\Delta\eta_{\W j}$, which then
should be enhanced for large $\Delta\eta_{\W j}$ in \ariadne as
compared to \pythia. This is also the case as shown in figure
\ref{fig:etaWj}b. Of course, the exact tree-level matrix element will
also contain contributions such as the one in \ref{fig:etaWj}a, and we
see that the enhancement at large $\Delta\eta_{\W j}$ is even more
significant, when ME corrections are added to \ariadne. It would be
interesting to see if including CKKW corrections also in \pythia would
bring it closer to \ariadne and ME4PS. This could be expected since
including higher order corrections could in the end make things more
insensitive to the particular kind of resummation used.

One of the advantages of the matrix element corrections is that
correlations between hard partons are more accurately described. This
may be important when eg.\ estimating backgrounds to different searches.
We will here consider the background to top production at the Tevatron
for the semi-leptonic channel which corresponds to \W+4-jets. In a
realistic top search one would use identified b-jets, but since our
\madgraph events do not include b-quarks we look at \W+4-jets in
general.

\FIGURE[t]{
      \epsfig{file=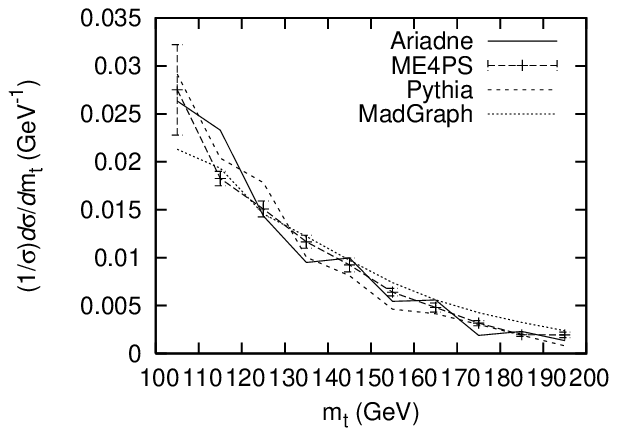,width=10cm}
  \caption{\label{fig:mt}
    This is the background for a search for the top quark.  A
    normalized distribution as a function of the mass of the top quark
    for \protect\ariadne (full line), \protect\pythia (short-dashed
    line), and ME4PS with 12~GeV cutoff (long-dashed with error bars).
    Also shown is the results from the pure tree-level matrix element
    without parton showers added using a 12~GeV cutoff (dotted
    line).}}

In figure \ref{fig:mt} we show the \W+4-jets background to the
top-mass distribution. We obtained it by using the \kT-algorithm to
cluster four jets and required that the jet scale was above 12~GeV.
Form these we found the two jets $j_1$ and $j_2$ with an invariant
mass $m_{12}$ closest to the \W\ mass. If no jets with
$|m_{12}-m_\W|<20$~GeV were found, the event was rejected. Then we
selected a third jet, $j_3$, so that the difference
$|m_{123}-m_{\W4}|$ was minimized, where $m_{123}$ is the invariant
mass of jets 1, 2 and 3, and $m_{\W4}$ is the invariant mass of the
\Wp and jet 4. If the difference $|m_{123}-m_{\W4}|<20$~GeV then the
event was accepted and the constructed top mass is defined as the
average of $m_{123}$ and $m_{\W4}$. Clearly the total cross section
would be underestimated by the leading-order predictions of \pythia
and \ariadne. In figure \ref{fig:mt} we therefore only show the
normalized shape and find that \pythia and \ariadne are quite similar
and that no significant change is introduced by the matrix element
correction to \ariadne. We also show the result from using the
tree-level 4-jet matrix elements directly, without reweighting and
adding a cascade, and find no large difference with the parton shower
approaches. Possibly the fall-off with $m_t$ can be said to be somewhat
weaker for the pure matrix elements.

\FIGURE[t]{
      \epsfig{file=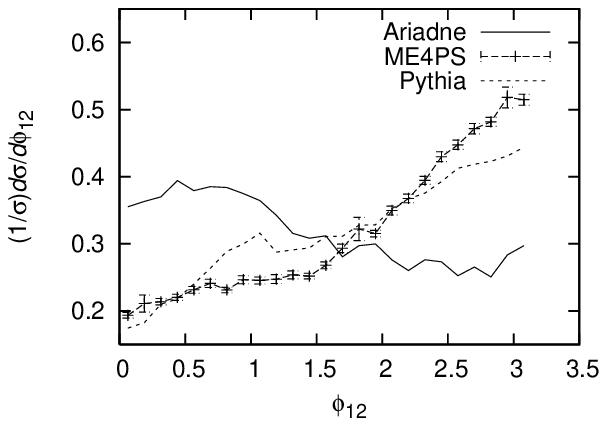,width=10cm}
  \caption{\label{fig:phi12}
    Normalized distribution of the azimuthal angle difference between
    the two hardest jets defined using the \kT-algorithm with a 12~GeV
    cutoff. The distribution is
    shown with \protect\ariadne (full line), \protect\pythia
    (short-dashed line), and ME4PS with 12~GeV cutoff (dashed line
    with error bars).}}

To focus more specifically on angular correlations, we finally look at
the azimuthal angle between the two hardest jets, $\phi_{12}$. This
observable is important for understanding how higher order emissions
influence the transverse momentum of the \W, \pTi{\W}. For
$\phi_{12}\sim\pi$, the emission of a second jet decreases the
\pTi{\W}, while for $\phi_{12}\sim0$ the \pTi{\W} is increased.  In
\ariadne, after a first emission of a gluon, a second gluon will be
radiated isotropically in azimuth in the rest system of the radiating
dipole. Since this dipole is boosted in the direction of the first
gluon, we expect that the second gluon is more likely to go in the
same direction. This is not true for \pythia, where successive
initial-state emissions are uncorrelated in azimuth (with the recoils,
the net effect is a bias towards large $\phi_{12}$). In figure
\ref{fig:phi12} we see the (normalized) $\phi_{12}$ distributions, and
indeed we find that \ariadne is more biased towards $\phi_{12}\sim0$.
Adding matrix element corrections removes this bias and brings the
distribution closer to the \pythia result. Hence this indicates that
the azimuthal correlations in standard \ariadne are not very well
modeled.

\section{Conclusions}
\label{sec:conclusions}

We have presented a way to implement a CKKW procedure for combining
events generated according to tree-level matrix elements for
W-production with the dipole cascade of \ariadne. Although the basic
principles are fairly simple, the details of our procedure is rather
involved, which is mandated by our aim to become as insensitive as
possible to the cutoff needed in the matrix element generation.

Our strategy is to take any partonic state generated by a MEG and try
to find a likely history of emissions which \ariadne would have
performed in order to generate this state. Rather than just using the
constructed emission scales to calculate analytic Sudakov form
factors to reweight the states, as is done in the original CKKW
procedure, we find exactly the Sudakov form factors \ariadne would
have used. In addition we reweight the states with the parton densities
functions and the so-called soft suppression function which \ariadne
would have used.

The PDF reweighting means that the overall normalization of cross
sections are still given by the leading order diagrams used by
standard \ariadne. However, we expect much improvement of the shapes 
of final state distributions as compared to the standard parton
shower description. In \cite{Krauss:2004bs} it was found that the CKKW
procedure reproduces well the shapes of distributions obtained by a
NLO program such as MC@NLO \cite{Frixione:2002ik}, although the
overall normalization needed to be adjusted with a K-factor. Although
not explicitly checked in this report, we expect that this will also
hold for our implementation.

We have presented several investigations into how the \ariadne program
is improved by adding matrix element corrections. In some cases we
also compared to the \pythia parton shower to get some insight into how
well these standard cascade programs reproduces higher order matrix
elements.

In one case we looked at the azimuthal correlation between the two
hardest jets, and found that the difference between \ariadne and
\pythia was large. When corrected with matrix elements, \ariadne came
much closer to \pythia, indicating that such azimuthal correlations
are not handled very well in standard \ariadne.

In our quasi-realistic top-background observable we found that
\ariadne and \pythia were quite close and that no drastic effect was
obtained by including matrix element corrections.

For the W--jet rapidity correlation we again found clear differences
between \ariadne and \pythia, and that these were even enhanced when
correcting \ariadne with matrix elements. This indicates that such
correlations are not very well described by \pythia, while standard
\ariadne does a better job, although it can be improved.

We believe that the rapidity correlations indicate that non-ordered
evolution is of importance for W-production at the Tevatron. Such
evolution is expected to be important in small-$x$ processes, and the
fact that it shows up here, where $x\sim m_W/\sqrt{S}\approx0.04$, may
be somewhat surprising. We also believe that the inclusion of
non-ordered evolution is why \ariadne is able to reproduce
experimental data on the small-\pT-distribution of the W and Z$^0$,
distributions which can only be described by \pythia if an
uncomfortably large intrinsic transverse momentum is added.

The fact that matrix element corrections can give us hints about where
unordered evolutions may become important, is an indication that it
would be very interesting to implement CKKW also for
DIS\footnote{Preliminary results for DIS have already been presented
  in \cite{abergthesis}.}  and compare with HERA data.  Also, at the
LHC where W-production may be argued to be a true small-$x$ process
($x\sim m_W/\sqrt{S}\ltaeq0.006$), it should be interesting to study
matrix element corrections. In fact also Higgs production at the LHC
may be considered to be a small-$x$ process. We will come back to these
processes in future publications.

%
%
%
%
%
%
%
%
%

\section*{Acknowledgments}

We would like to thank Stephen Mrenna, Peter Richardson, Torbjörn
Sjöstrand and Christoffer Åberg for useful discussions. Special thanks
to Stephen Mrenna for supplying us with event files generated with
\madgraph/\madevent.

\bibliographystyle{utcaps}  
\bibliography{references} 

\providecommand{\href}[2]{#2}\begingroup\raggedright\begin{thebibliography}{10%
}\itemsep 0mm

\bibitem{Catani:2001cc}
S.~Catani, F.~Krauss, R.~Kuhn, and B.~R. Webber {\em JHEP} {\bf 11} (2001) 063,
\href{http://arXiv.org/abs/hep-ph/0109231}{{\tt hep-ph/0109231}}.

\bibitem{Lonnblad:2001iq}
L.~Lönnblad {\em JHEP} {\bf 05} (2002) 046,
\href{http://arXiv.org/abs/hep-ph/0112284}{{\tt hep-ph/0112284}}.

\bibitem{Gustafson:1988rq}
G.~Gustafson and U.~Pettersson {\em Nucl.~Phys.} {\bf B306} (1988)
746.

\bibitem{Gustafson:1986db}
G.~Gustafson {\em Phys.~Lett.} {\bf B175} (1986)
453.

\bibitem{Lonnblad:1992tz}
L.~Lönnblad {\em Comput.~Phys.~Commun.} {\bf 71} (1992)
15--31.

\bibitem{Krauss:2002up}
F.~Krauss {\em JHEP} {\bf 08} (2002) 015,
\href{http://www.arXiv.org/abs/hep-ph/0205283}{{\tt hep-ph/0205283}}.

\bibitem{Krauss:2004bs}
F.~Krauss, A.~Schalicke, S.~Schumann, and G.~Soff {\em Phys. Rev.} {\bf D70}
  (2004) 114009,
\href{http://www.arXiv.org/abs/hep-ph/0409106}{{\tt hep-ph/0409106}}.

\bibitem{Krauss:2005nu}
F.~Krauss, A.~Schaelicke, S.~Schumann, and G.~Soff
\href{http://www.arXiv.org/abs/hep-ph/0503280}{{\tt hep-ph/0503280}}.

\bibitem{Mrenna:2003if}
S.~Mrenna and P.~Richardson {\em JHEP} {\bf 05} (2004) 040,
\href{http://www.arXiv.org/abs/hep-ph/0312274}{{\tt hep-ph/0312274}}.

\bibitem{MLM}
M.~Mangano, ``The so-called MLM prescription for ME/PS matching.''
  \texttt{http://www-cpd.fnal.gov/personal/mrenna/tuning/nov2002/mlm.pdf}.
\newblock Talk presented at the Fermilab ME/MC Tuning Workshop, October 4,
  2002.

\bibitem{Andersson:1989gp}
B.~Andersson, G.~Gustafson, L.~L{\"o}nnblad, and U.~Pettersson {\em Z. Phys.}
  {\bf C43} (1989)
625.

\bibitem{Lonnblad:1996ex}
L.~Lönnblad {\em Nucl.~Phys.} {\bf B458} (1996) 215--230,
\href{http://arXiv.org/abs/hep-ph/9508261}{{\tt hep-ph/9508261}}.

\bibitem{Sjostrand:2001yu}
{T.~Sj\"ostrand, L.~L\"onnblad, and S.~Mrenna}, ``PYTHIA 6.2: Physics and
  manual,''
\href{http://www.arXiv.org/abs/arXiv:hep-ph/0108264}{{\tt
  arXiv:hep-ph/0108264}}.

\bibitem{Bengtsson:1987rw}
M.~Bengtsson and T.~Sjöstrand {\em Z. Phys.} {\bf C37} (1988)
465.

\bibitem{Corcella:2000bw}
G.~Corcella {\em et al.} {\em JHEP} {\bf 01} (2001) 010,
\href{http://arXiv.org/abs/hep-ph/0011363}{{\tt hep-ph/0011363}}.

\bibitem{Gleisberg:2003xi}
T.~Gleisberg {\em et al.} {\em JHEP} {\bf 02} (2004) 056,
\href{http://www.arXiv.org/abs/hep-ph/0311263}{{\tt hep-ph/0311263}}.

\bibitem{Krauss:2005re}
F.~Krauss, A.~Schalicke, and G.~Soff
\href{http://www.arXiv.org/abs/hep-ph/0503087}{{\tt hep-ph/0503087}}.

\bibitem{Gribov:1972ri}
V.~N. Gribov and L.~N. Lipatov {\em Yad. Fiz.} {\bf 15} (1972)
781--807.

\bibitem{Lipatov:1975qm}
L.~N. Lipatov {\em Sov. J. Nucl. Phys.} {\bf 20} (1975)
94--102.

\bibitem{Altarelli:1977zs}
G.~Altarelli and G.~Parisi {\em Nucl. Phys.} {\bf B126} (1977)
298.

\bibitem{Dokshitzer:1977sg}
Y.~L. Dokshitzer {\em Sov. Phys. JETP} {\bf 46} (1977)
641--653.

\bibitem{Kuraev:1976ge}
E.~A. Kuraev, L.~N. Lipatov, and V.~S. Fadin {\em Sov. Phys. JETP} {\bf 44}
  (1976)
443--450.

\bibitem{Kuraev:1977fs}
E.~A. Kuraev, L.~N. Lipatov, and V.~S. Fadin {\em Sov. Phys. JETP} {\bf 45}
  (1977)
199--204.

\bibitem{Balitsky:1978ic}
I.~I. Balitsky and L.~N. Lipatov {\em Sov. J. Nucl. Phys.} {\bf 28} (1978)
822--829.

\bibitem{Ciafaloni:1988ur}
M.~Ciafaloni {\em Nucl. Phys.} {\bf B296} (1988)
49.

\bibitem{Catani:1990yc}
S.~Catani, F.~Fiorani, and G.~Marchesini {\em Phys. Lett.} {\bf B234} (1990)
339.

\bibitem{Catani:1990sg}
S.~Catani, F.~Fiorani, and G.~Marchesini {\em Nucl. Phys.} {\bf B336} (1990)
18.

\bibitem{Marchesini:1995wr}
G.~Marchesini {\em Nucl. Phys.} {\bf B445} (1995) 49--80,
\href{http://arXiv.org/abs/hep-ph/9412327}{{\tt hep-ph/9412327}}.

\bibitem{Bengtsson:1987hr}
M.~Bengtsson and T.~Sjostrand {\em Phys.~Lett.} {\bf B185} (1987)
435.

\bibitem{Bengtsson:1987et}
M.~Bengtsson and T.~Sjöstrand {\em Nucl.~Phys.} {\bf B289} (1987)
810.

\bibitem{Seymour:1995we}
M.~H. Seymour {\em Nucl.~Phys.} {\bf B436} (1995) 443--460,
\href{http://arXiv.org/abs/hep-ph/9410244}{{\tt hep-ph/9410244}}.

\bibitem{Seymour:1995df}
M.~H. Seymour {\em Comp.~Phys.~Commun.} {\bf 90} (1995) 95--101,
\href{http://arXiv.org/abs/hep-ph/9410414}{{\tt hep-ph/9410414}}.

\bibitem{Seymour:1994ti}
M.~H. Seymour, ``Matrix element corrections to parton shower simulation of deep
  inelastic scattering.''
\newblock Contributed to 27th International Conference on High Energy Physics
  (ICHEP), Glasgow, Scotland, 20-27 Jul 1994.

\bibitem{Lonnblad:1995wk}
L.~Lönnblad {\em Z.~Phys.} {\bf C65} (1995)
285--292.

\bibitem{Miu:1998ju}
G.~Miu and T.~Sjöstrand {\em Phys.~Lett.} {\bf B449} (1999) 313--320,
\href{http://arXiv.org/abs/hep-ph/9812455}{{\tt hep-ph/9812455}}.

\bibitem{Mrenna:1999mq}
S.~Mrenna, ``Higher order corrections to parton showering from resummation
  calculations,''
\href{http://arXiv.org/abs/hep-ph/9902471}{{\tt hep-ph/9902471}}.

\bibitem{Corcella:1999gs}
G.~Corcella and M.~H. Seymour {\em Nucl.~Phys.} {\bf B565} (2000) 227--244,
\href{http://arXiv.org/abs/hep-ph/9908388}{{\tt hep-ph/9908388}}.

\bibitem{kTalgorithm}
Y.~L. Dokshitzer. in \textit{Workshop on Jet studies at LEP and HERA}, Durham
  1990, see \textit{J.~Phys.} \textbf{G17} (1991) 1572ff.

\bibitem{Catani:1991hj}
S.~Catani, Y.~L. Dokshitzer, M.~Olsson, G.~Turnock, and B.~R. Webber {\em Phys.
  Lett.} {\bf B269} (1991)
432--438.

\bibitem{Lonnblad:1993qd}
L.~Lönnblad {\em Z.~Phys.} {\bf C58} (1993)
471--478.

\bibitem{Moretti:1998qx}
S.~Moretti, L.~Lönnblad, and T.~Sjöstrand {\em JHEP} {\bf 08} (1998) 001,
\href{http://arXiv.org/abs/hep-ph/9804296}{{\tt hep-ph/9804296}}.

\bibitem{Andersson:1990ki}
B.~Andersson, G.~Gustafson, and L.~L{\"o}nnblad {\em Nucl. Phys.} {\bf B339}
  (1990)
393--406.

\bibitem{Hamacher:1995df}
K.~Hamacher and M.~Weierstall
\href{http://www.arXiv.org/abs/hep-ex/9511011}{{\tt hep-ex/9511011}}.

\bibitem{Brook:1995nn}
N.~Brook, R.~G. Waugh, T.~Carli, R.~Mohr, and M.~Sutton. Prepared for Workshop
  on Future Physics at HERA (Preceded by meetings 25-26 Sep 1995 and 7-9 Feb
  1996 at DESY), Hamburg, Germany, 30-31 May 1996.

\bibitem{Boos:2001cv}
E.~Boos {\em et al.}
\href{http://arXiv.org/abs/hep-ph/0109068}{{\tt hep-ph/0109068}}.

\bibitem{Schaelicke:2005nv}
A.~Schaelicke and F.~Krauss
\href{http://www.arXiv.org/abs/hep-ph/0503281}{{\tt hep-ph/0503281}}.

\bibitem{Catani:1993hr}
S.~Catani, Y.~L. Dokshitzer, M.~H. Seymour, and B.~R. Webber {\em Nucl. Phys.}
  {\bf B406} (1993)
187--224.

\bibitem{nilsprep}
N.~Lavesson, L.~Lönnblad, and C.~Åberg. Preprint in preparation.

\bibitem{Maltoni:2002qb}
F.~Maltoni and T.~Stelzer {\em JHEP} {\bf 02} (2003) 027,
\href{http://www.arXiv.org/abs/hep-ph/0208156}{{\tt hep-ph/0208156}}.

\bibitem{Pumplin:2002vw}
J.~Pumplin {\em et al.} {\em JHEP} {\bf 07} (2002) 012,
\href{http://www.arXiv.org/abs/hep-ph/0201195}{{\tt hep-ph/0201195}}.

\bibitem{Thome:2004sk}
E.~Thome
\href{http://www.arXiv.org/abs/hep-ph/0401121}{{\tt hep-ph/0401121}}.

\bibitem{Frixione:2002ik}
S.~Frixione and B.~R. Webber
\href{http://arXiv.org/abs/hep-ph/0204244}{{\tt hep-ph/0204244}}.

\bibitem{abergthesis}
C.~Åberg, ``Correcting the Colour Dipole Cascade with Fixed Order Matrix
  Elements in Deep Inelastic Scattering.'' Diploma thesis, LU-TP 04-25.

\end{thebibliography}\endgroup

\appendix

\section{Appendix: Construction}
\label{sec:appendix}

Here we describe in some detail the different kinds of steps possible
when constructing intermediate partonic states from events generated
by a MEG. In figure \ref{fig:reconstuctions} the different steps are
shown schematically.

\FIGURE[t]{
  \begin{picture}(360,210)(0,0)
    \Text(10,200)[]{(a)}
    \LongArrow(85,175)(105,175)

    \Line(10,180)(55,200)
    \Line(10,170)(55,150)
    \Gluon(10,175)(60,175){3}{6}
    \DashCArc(50,187)(15,-30,50){2}
    \DashCArc(50,162)(15,-50,30){2}
    \Text(70,200)[]{\scriptsize $p_3$}
    \Text(70,175)[]{\scriptsize $\g_2$}
    \Text(70,150)[]{\scriptsize $p_1$}

    \Line(115,180)(160,195)
    \Line(115,170)(160,155)
    \DashCArc(145,175)(25,-40,40){2}
    \Text(170,195)[]{\scriptsize $p_3$}
    \Text(170,155)[]{\scriptsize $p_1$}

    \Text(10,130)[]{(b)}
    \LongArrow(85,105)(105,105)

    \Line(10,110)(55,130)
    \Line(10,100)(55,80)
    \Line(10,105)(60,105)
    \DashCArc(50,92)(15,-50,30){2}
    \Text(70,130)[]{\scriptsize $\qbar_3$}
    \Text(70,105)[]{\scriptsize $\q_2$}
    \Text(70,80)[]{\scriptsize $p_1$}

    \Gluon(115,110)(160,125){3}{6}
    \Line(115,100)(160,85)
    \DashCArc(145,105)(25,-40,40){2}
    \Text(170,125)[]{\scriptsize $\g$}
    \Text(170,85)[]{\scriptsize $p_1$}

    \Text(10,60)[]{(c)}
    \LongArrow(85,35)(105,35)

    \Line(10,40)(55,60)
    \Line(10,30)(55,10)
    \Line(10,27)(55,7)
    \Oval(55,8.5)(5,3)(0)
    \Gluon(10,35)(60,35){3}{6}
    \DashCArc(50,47)(15,-30,50){2}
    \DashCArc(50,22)(15,-50,30){2}
    \Text(70,60)[]{\scriptsize $p_1$}
    \Text(70,35)[]{\scriptsize $\g_2$}
    \Text(70,10)[]{\scriptsize $r_3$}

    \Line(115,40)(160,55)
    \Line(115,30)(160,15)
    \Line(115,27)(160,12)
    \Oval(115,28.5)(5,3)(0)
    \Oval(160,13.5)(5,3)(0)
    \DashCArc(145,35)(25,-40,40){2}
    \Text(170,55)[]{\scriptsize $p_1$}
    \Text(170,15)[]{\scriptsize $r_3$}

    \Text(210,200)[]{(d)}
    \LongArrow(285,175)(305,175)

    \Line(210,180)(255,200)
    \Line(210,183)(255,203)
    \Oval(255,201.5)(5,3)(0)
    \Line(210,170)(255,150)
    \Line(210,167)(255,147)
    \Oval(255,148.5)(5,3)(0)
    \Gluon(210,175)(260,175){3}{6}
    \DashCArc(250,187)(15,-30,50){2}
    \DashCArc(250,162)(15,-50,30){2}
    \Text(270,200)[]{\scriptsize $r_1$}
    \Text(270,175)[]{\scriptsize $\g_2$}
    \Text(270,150)[]{\scriptsize $r_3$}

    \Line(315,180)(360,195)
    \Line(315,183)(360,198)
    \Oval(360,197.5)(5,3)(0)
    \Line(315,170)(360,155)
    \Line(315,167)(360,152)
    \Oval(360,153.5)(5,3)(0)
    \DashCArc(345,175)(25,-40,40){2}
    \Text(370,155)[]{\scriptsize $r_1$}
    \Text(370,195)[]{\scriptsize $r_3$}

    \Text(210,130)[]{(e)}
    \LongArrow(285,105)(305,105)

    \Line(210,113)(255,133)
    \Line(210,98)(260,98)
    \Oval(260,98)(3,3)(0)
    \Line(210,108)(260,108)
    \Line(210,105)(260,105)
    \Oval(260,106.5)(5,3)(0)
    \DashCArc(247,120)(15,-30,50){2}
    \DashCArc(247,85)(15,-50,30){2}
    \Text(270,130)[]{\scriptsize $\q_2$}
    \Text(270,100)[]{\scriptsize $r_3'$}
    \Text(270,110)[]{\scriptsize $r_3$}
    \Text(270,75)[]{\scriptsize $H_1$}

    \Line(315,111)(360,111)
    \Line(315,108)(360,108)
    \Line(315,105)(360,105)
    \GOval(360,108)(7,3)(0){0}
    \Line(315,98)(360,98)
    \Oval(360,98)(3,3)(0)
    \DashCArc(347,85)(15,-50,30){2}
    \Text(370,100)[]{\scriptsize $r_3'$}
    \Text(370,110)[]{\scriptsize $h_3$}
    \Text(370,75)[]{\scriptsize $H_1$}

    \Text(210,60)[]{(f)}
    \LongArrow(285,25)(305,25)

    \Line(210,35)(255,43)
    \Line(210,14)(260,14)
    \Oval(260,14)(3,3)(0)
    \Line(210,28)(260,28)
    \Line(210,22)(260,22)
    \Line(210,25)(260,25)
    \GOval(260,25)(7,3)(0){0}
    \DashCArc(247,5)(15,-50,30){2}
    \DashCArc(240,47)(20,-15,20){2}
    \Text(270,40)[]{\scriptsize $\q_2$}
    \Text(270,13)[]{\scriptsize $r_3$}
    \Text(270,23)[]{\scriptsize $h_3$}
    \Text(270,-5)[]{\scriptsize $H_1$}
    \Text(270,55)[]{\scriptsize $H'_1$}

    \Line(315,22)(360,22)
    \Line(315,25)(360,25)
    \Oval(360,23.5)(5,3)(0)
    \Line(315,14)(360,14)
    \Oval(360,14)(3,3)(0)
    \DashCArc(347,5)(15,-50,30){2}
    \Text(370,14)[]{\scriptsize $r_3$}
    \Text(370,24)[]{\scriptsize $r_3'$}
    \Text(370,-5)[]{\scriptsize $H_1$}
    \Text(370,52)[]{\scriptsize $H'_1$}
    \DashCArc(340,35)(20,-15,35){2}

  \end{picture}
  \caption{\label{fig:reconstuctions}Symbolic pictures of possible steps
    which are possible when constructing possible intermediate
    partonic states from events generated by a MEG. The different
    possibilities are described in the text. The dashed lines indicate
    colour-connections.}}

\begin{anumerate}
\item $p_1\dash\g_2\dash p_3\longrightarrow p_1\dash p_3$: A gluon,
  colour-connected to two non-remnant partons, $p_1$ and $p_3$, is
  constructed to a single dipole between $p_1$ and $p_3$. The
  splitting function is given by one of \eqsref{eq:dipsplitqq} --
  (\ref{eq:dipsplitgg}) depending on whether $p_1$ and $p_3$ are
  gluons or quarks. The scale is given by the invariant \pT\ in
  \eqref{eq:invpt}. No PDF ratio is relevant. In the rest frame of the
  construction, $p_1$ will retain its direction if it is a gluon and
  $p_3$ is a quark, and vice versa. If both $p_1$ and $p_3$ are gluons
  or both are quarks, the direction of $p_1$ is rotated away from the
  original $p_3$ direction with an angle $\beta x_1^2/(x_1^2+x_3^2)$,
  where $\beta$ is the original angle between $p_1$ and $p_3$. This
  corresponds to the standard recoil treatment for gluon emission in
  \ariadne.
\item $p_1\dash\q_2\,\,\qbar_3\longrightarrow p_1\dash\g$. A
  \q\qbar-pair is constructed into a gluon as long as they are each
  others anti-particles, are connected to different strings, and the
  end-points of these strings are not remnants of the same incoming
  hadron. The splitting function is given by \eqref{eq:dipsplitQQ}.
  The scale is given by the invariant \pT\ in \eqref{eq:invpt}. No PDF
  ratio is relevant.  In the rest system of the construction, $p_1$
  will retain its direction.
\item $p_1\dash\g_2\dash r_3\longrightarrow p_1\dash r_3$: A gluon,
  colour-connected to one remnant, $r_3$ and one non-remnant parton,
  $p_1$, is constructed to a single dipole between $p_1$ and $r_3$.
  The splitting function is given by \eqsref{eq:dipsplitqq} or
  (\ref{eq:dipsplitqg}) depending on whether $p_1$ is a quark or a
  gluon. 
  If $r_3$ is one of two remnants of the same hadron (this corresponds
  to an extracted gluon), the PDF ratio is taken to be $\Theta$,
  otherwise it is $\Theta/z$. If a \W\ is present in the event and it
  is \emph{close} to $\g_2$, the transverse momentum of the gluon in
  the center of mass system of $p_1$ and $r_3$ is given to the \W, and
  the longitudinal momentum is absorbed by $p_1$ and $r_3$. The scale
  is given by the invariant \pT\ in \eqref{eq:invpt} (calculated as if
  no \W\ was close, ie.\ the transverse momenta of the gluon is
  transfered to $p_1$ and $r_3$ with the weight
  $x_i^2/(x_1^2+x^2_3)$).  Here \emph{close} means that
  $p_{+g}<p_{+\W}$ and $p_{-g}<p_{-\W}$, where $p_{\pm W}$ is
  calculated for the constructed W momenta. If there is no \W\ close
  by the momentum of the gluon is shared by $p_1$ and $r_3$, where
  $r_3$ retains its direction. The scale is given by the invariant \pT
  .
\item $r_1\dash\g_2\dash r_3\longrightarrow r_1\dash r_3$: A gluon
  connected to two remnants, one from each incoming hadron, is
  constructed to a single dipole between the remnants. The splitting
  function is given by \eqsref{eq:dipsplitqq}. When calculating the scale
  a fraction $x_i^2/(x_1^2+x^2_3)$ of the transverse momenta from the 
  gluon is transfered to each of the remnants and the scale is given
  by the invariant \pT\ in \eqref{eq:invpt}. The PDF ratio is given by
  the product of the $\Theta$ on each side, divided by $z$ if the
  corresponding remnant is not one of two remnants of the same hadron.
  The transverse momentum of the gluon is transfered to the \W\ if one
  is present, otherwise it is transfered to the hard subsystem
  containing the rest of the non-remnant partons in the event. The
  longitudinal momentum is divided between $r_1$ and $r_3$.
\item \label{item:e} $H_1\,\,r_3'\,\,\q_2\dash r_3\longrightarrow
  H_1\,\, r_3'\,\,h_3$: This corresponds to the inverse of an
  initial-sate $\g\to\q$ splitting. For a quark, $\q_2$, connected to
  a remnant, $r_3$, and a hard subsystem, $H_1$ (which contains the
  \W\ if present), connected to another remnant, $r_3'$, from the same
  incoming hadron and arising from the extraction of a corresponding
  anti-quark, $\qbar'$, a hadron, $h_3$, is formed from $\q_2$ and
  $r_3$. The splitting function is the standard Altarelli--Parisi one,
  $P_{\g\to\q}(z)$. The scale is the squared transverse momentum of
  $\q_2$ in the rest frame of the event. The PDF ratio is the same as
  would have been used in a conventional parton shower. The transverse
  momentum of $\q_2$ is transfered to $H_1$, and the longitudinal
  momentum is shared between $H_1$, $r_3'$ and $h_3$.  The relative
  sharing of longitudinal momenta between $r_3'$ and $h_3$ is the same
  as for the original $r_3'$ and $r_3$.
\item \label{item:f} $H_1\,\,\q_2\,\, r_3\,(h_3)\longrightarrow
  H_1\,\,r_3\,\,r_3'$: A quark, $\q_2$, which may have been extracted
  from a hadron resulting in a remnant $r_3$ may be absorbed into a
  the remnant, constructing an initial-state $\q\to\g$ splitting.
  The remnant is split into two, possibly together with a remnant
  hadron, $h_3$, if $\q_2$ was a sea-quark. The splitting function is
  the standard Altarelli--Parisi one, $P_{\q\to\g}(z)$. The scale is
  the squared transverse momentum of $\q_2$ in the rest frame of the
  event. The PDF ratio is the same as would have been used in a
  conventional parton shower. The transverse momentum of $\q_2$ is
  transfered to the spectator hard subsystem, $H_1$, and the
  longitudinal momentum is shared between $H_1$, $r_3$ and $r_3'$. The
  relative sharing of longitudinal momenta between $r_3$ and $r_3'$ is
  the same as for the original $r_3$ and $h_3$ if $h_3$ was present,
  otherwise the momenta is shared as is normally done in \ariadne when
  a gluon is extracted from a hadron. Note that there is no
  corresponding emission in \ariadne.
\end{anumerate}

\end{document}